\global\def\draftcontrol{0}

   \def\versionno{ higher-derivative-transport}

\catcode`\@=11

\expandafter\ifx\csname draftcontrol\endcsname\relax\global\def\draftcontrol{0}
\fi

{\count255=\time\divide\count255 by 60
\xdef\hourmin{\number\count255}
\multiply\count255 by-60\advance\count255 by\time
\xdef\hourmin{\hourmin:\ifnum\count255<10 0\fi\the\count255}}
\def\draftdate{\number\month/\number\day/\number\year\ \ \ \hourmin }

\newcommand\makepapertitle{\par
  \begingroup
    \renewcommand\thefootnote{\@fnsymbol\c@footnote}%
    \def\@makefnmark{\rlap{\@textsuperscript{\normalfont\@thefnmark}}}%
    \long\def\@makefntext##1{\parindent 1em\noindent
            \hb@xt@1.8em{%
                \hss\@textsuperscript{\normalfont\@thefnmark}}##1}%
     \newpage
     \global\@topnum\z@   
     \@makepapertitle
     \thispagestyle{empty}\@thanks
  \endgroup
  \setcounter{footnote}{0}%
  \global\let\thanks\relax
  \global\let\makepapertitle\relax
  \global\let\@makepapertitle\relax
  \global\let\@thanks\@empty
  \global\let\@author\@empty
  \global\let\@date\@empty
  \global\let\@title\@empty
  \global\let\title\relax
  \global\let\author\relax
  \global\let\date\relax
  \global\let\and\relax
  \def\version{\let\version\@version\@gobble}
}
\def\@makepapertitle{%
  \newpage
   \ifnum\draftcontrol=1 {}
   \version\versionno
   \vskip 3em%
   \else
   \hfill\hbox to 3cm {\parbox{4cm}{\@pubnum}\hss}%
   \vskip 3em%
   \fi
   \begin{center}%
   \let \footnote \thanks
     {\LARGE {\@title}}%
     \vskip 1.5em%
     {\normalsize
       \lineskip .5em%
       \begin{tabular}[t]{c}%
         \@author
       \end{tabular}\par}%
     \vskip 1.5em%
     {\@bstract}%
     \end{center}%
     \vskip 1.5em
     \@date%
   \par
}

\gdef\@pubnum{}
\def\pubnum#1{%
  \gdef\@pubnum{#1}}

\gdef\@bstract{}
\def\Abstract#1{%
  \gdef\@bstract{%
   \parbox{\textwidth-0pc}{%
   \centerline{\bf Abstract}\penalty1000%
\kern.2cm%
\noindent
\renewcommand\baselinestretch{1.0}%
{#1}}}
}

\def\ps@paper{\let\@mkboth\@gobbletwo%
     \ifnum\draftcontrol=1
    \def\@oddfoot{\hbox to \textwidth{\tiny \versionno \hfil\tiny\draftdate}%
    \hskip -\textwidth \hbox to \textwidth{\hfil\rm\thepage\hfil}}%
     \else\def\@oddfoot{\hbox to \textwidth{\hfil\rm\thepage\hfil}}
     \fi
     \let\@evenfoot\@oddfoot
}

\def\body{\clearpage
          \pagestyle{paper}
    }

\def\@version#1{\ifnum\draftcontrol=1
\typeout{}\typeout{#1}\typeout{}
\vskip3mm\centerline{\hbox{\fbox{\normalsize{\tt DRAFT -- #1 -- }
                   {\draftdate}}}}\vskip3mm
\fi}
\let\version\@version
\long\def\eqlabel#1{\ifnum\draftcontrol=1
                    \tag@false  
                    \tag*{(\theequation) \hbox to -0.2cm{\hspace{0cm}\small{#1}\hss}}
                    \refstepcounter{equation}
                    \edef\@currentlabel{\theequation}
                    \ltx@label{#1}          
                    \else
                    \label{#1}
                    \fi
                    }
\let\st@bibitem\@bibitem
\let\st@lbibitem\@lbibitem
\ifnum\draftcontrol=1
  \def\@bibitem#1{%
    \st@bibitem{#1}\a@@label{#1}\ignorespaces}
  \def\@lbibitem[#1]#2{%
    \st@lbibitem[#1]{#2}\a@@label{#2}\ignorespaces}
  \def\a@@label#1{%
    \gdef\a@lab{\smash{\normalfont\small#1}}
    \ifvmode
      \if@inlabel
        \global\setbox\@labels\hbox{%
          \llap{\a@lab\let\a@lab\relax
                \kern\@totalleftmargin\kern\marginparsep}%
          \box\@labels}%
      \fi
    \fi}
\fi

\documentclass[12pt,letterpaper]{article}

\usepackage{amsmath,amssymb,array,calc,epsfig,rotating,bm}
\usepackage[sort]{cite}
\usepackage{graphicx}
\usepackage{psfrag,verbatim}
\usepackage{xcolor}
\usepackage{hyperref}

\ifnum\draftcontrol=1
\fi

\renewcommand\baselinestretch{1.25}
\renewcommand\section{\@startsection {section}{1}{\z@}%
                                   {-3.5ex \@plus -1ex \@minus -.2ex}%
                                   {2.3ex \@plus.2ex}%
                                   {\normalfont\large\bfseries}}
\renewcommand\subsection{\@startsection{subsection}{2}{\z@}%
                                   {-3.25ex\@plus -1ex \@minus -.2ex}%
                                   {1.5ex \@plus .2ex}%
                                   {\normalfont\normalsize\bfseries}}
\renewcommand\subsubsection{\@startsection{subsubsection}{3}{\z@}%
                                   {-3.25ex\@plus -1ex \@minus -.2ex}%
                                   {1.5ex \@plus .2ex}%
                                   {\normalfont\normalsize\it}}
\renewcommand\paragraph{\@startsection{paragraph}{4}{\z@}%
                                   {-3.25ex\@plus -1ex \@minus -.2ex}%
                                   {1.5ex \@plus .2ex}%
                                   {\normalfont\normalsize\bf}}

\numberwithin{equation}{section}

\def\revise#1       {\raisebox{-0em}{\rule{3pt}{1em}}%
                     \marginpar{\raisebox{.5em}{\vrule width3pt\
                     \vrule width0pt height 0pt depth0.5em
                     \hbox to 0cm{\hspace{0cm}{%
                     \parbox[t]{4em}{\raggedright\footnotesize{#1}}}\hss}}}}

\newcommand\nxt[1]  {\\\fnxt#1}
\newcommand{\ie}{{\it i.e.,}\ }
\newcommand{\eg}{{\it e.g.,}\ }

\def\cala         {{\cal A}}

\def\calb         {{\cal B}}
\def\calc         {{\cal C}}
\def\cald         {{\cal D}}
\def\cale         {{\cal E}}
\def\calf         {{\cal F}}

\def\cali         {{\cal I}}

\def\calk         {{\cal K}}
\def\call         {{\cal L}}
\def\calm         {{\cal M}}
\def\caln         {{\cal N}}
\def\calo         {{\cal O}}

\def\complex      {{\mathbb C}}

\def\del          {\partial}

\def\Im           {{\rm Im\hskip0.1em}}

\def\sqr#1#2{{\vcenter{\vbox{\hrule height.#2pt
 \hbox{\vrule width.#2pt height#1pt \kern#1pt
 \vrule width.#2pt}\hrule height.#2pt}}}}

\newcommand{\ft}[2]{{\textstyle{\frac{#1}{#2}}}}

\newcommand{\kk}{\mathfrak{q}}

\def\dd{\delta}

\def\aa1{\phi}
\def\cc1{\psi}

\newcommand{\hq}{\mathfrak{q}}
\newcommand{\hw}{\mathfrak{w}}

\catcode`\@=12

\begin{document}


\title{\bf Holographic transport beyond the supergravity approximation}

\date{February 16, 2024}

\author{
Alex Buchel$^{1,2}$\,,  Sera Cremonini$^{3}$ and Laura Early$^{3}$\\[0.4cm]
\it $^1$Department of Physics and Astronomy\\ 
\it University of Western Ontario\\
\it London, Ontario N6A 5B7, Canada\\
\it $^2$Perimeter Institute for Theoretical Physics\\
\it Waterloo, Ontario N2J 2W9, Canada\\
\it $^3$Department of Physics, Lehigh University\\
\it Bethlehem, PA, 18015, U.S.A.\\[0.4cm]\\
}

\Abstract{We set up a unified framework to efficiently compute the shear and
bulk viscosities of strongly coupled gauge theories with gravitational
holographic duals involving higher derivative corrections. We consider
both Weyl$^4$ corrections, encoding the finite 't Hooft coupling
corrections of the boundary theory, and Riemann$^2$ corrections,
responsible for non-equal central charges $c\ne a$ of the theory at
the ultraviolet fixed point.  Our expressions for the viscosities in
higher derivative holographic models are extracted from a radially
conserved current and depend only on the horizon data.
}

\makepapertitle

\body

\version\versionno
\tableofcontents

\section{Introduction and summary}\label{intro}

For well over two decades there has been an ongoing program to apply the holographic gauge/gravity duality to gain insights into 
the dynamics of strongly coupled quantum phases of matter \cite{Casalderrey-Solana:2011dxg}. 
Indeed, the techniques of holography have been adopted to probe the transport properties of a wide spectrum of strongly correlated systems, ranging from 
the QCD quark gluon plasma (QGP) to high temperature superconductors, strange metals and a variety of electronic materials.  
Within this program,  the most elegant result to date remains the universality \cite{Policastro:2001yc,Buchel:2003tz} of the shear viscosity $\eta$ to entropy $s$ ratio, 
\begin{equation}
\label{universal}
\frac{\eta}{s} = \frac{\hbar}{4\pi k_B} \, ,
\end{equation}
which holds in strongly coupled gauge theories in the limit of an infinite number of colors, $N \rightarrow \infty$, and
infinite 't Hooft coupling, $\lambda \rightarrow \infty$. These theories are dual to Einstein gravity coupled to an arbitrary matter sector, with the result (\ref{universal}) relying on the additional assumption that rotational invariance is preserved\footnote{For a recent discussion of $\eta/s$ in anisotropic theories we refer the reader to \cite{Baggioli:2023yvc}.}.

The significance of (\ref{universal}) can be traced not only to its universal nature, but also to the fact that its value is remarkably close to the experimental range extracted from the QGP data at RHIC and at the LHC. 
This led to the compelling KSS proposal \cite{Kovtun:2003wp,Kovtun:2004de} 
that the shear viscosity might obey 
a fundamental lower bound in nature, $ \frac{\eta}{s}  \geq \frac{1}{4\pi}$ (from now on we take $\hbar=k_B=1$).
Despite its appeal, it is now well understood that the KSS bound can be violated in a number of ways, either by relaxing symmetries or by introducing higher derivative \emph{curvature} corrections to the low-energy gravitational action (see \emph{i.e.} \cite{Cremonini:2011iq} for a review). 
Indeed, notable early examples of the effects of higher derivatives include 
Weyl$^4$ corrections to Einstein gravity \cite{Buchel:2004di}, which encode
finite $\lambda$
effects in the dual gauge theory, and 
Riemann$^2$ corrections \cite{Kats:2007mq,Brigante:2008gz}, which describe finite $N$ effects in the dual theory\footnote{These are due to non-equal central
charges $c\ne a$ of the gauge theory at the UV fixed point.}.
Holographic models involving a more complicated matter sector have been used to study the temperature dependence of $\eta/s$ \cite{Buchel:2010wf,Cremonini:2012ny} and have direct applications to the QGP, where the temperature variations of $\eta$ are expected to play an important role (see \emph{e.g.} the recent review \cite{Rougemont:2023gfz}).
A key lesson that has emerged from these studies is that the universality of $\eta/s$ is generically lost once we move away from the limits $\lambda,N \rightarrow \infty$ (or appropriately break symmetries). Moreover, other transport coefficients have failed to exhibit the simple universal behavior encoded in (\ref{universal}). 
 
In addition to $\eta$, the bulk viscosity $\zeta$ has also attracted considerable attention in 
holography (see e.g. \cite{Parnachev:2005hh,Benincasa:2005iv,Buchel:2005cv,Buchel:2007mf,Gubser:2008sz,Buchel:2008uu,Gursoy:2009kk,Buchel:2011wx,Buchel:2011uj,Eling:2011ms,Buchel:2011yv}), largely because of its relevance to the physics of the QGP near the deconfinement transition, where $\zeta$ is expected to rise dramatically. As non-zero bulk viscosity requires theories with broken conformal symmetry, holographic model building has typically involved adding bulk scalars with non-trivial profiles. 
Since the latter also yields a non-trivial temperature dependence for $\eta/s$, such
holographic models have played a prominent role in the attempts to 
build realistic models of QCD \cite{Gursoy:2007cb,Gursoy:2007er}.

In holography, transport coefficients such as $\eta$ and $\zeta$ can be extracted in a number of complementary ways, e.g. computing correlators of the stress energy tensor and using Kubo formulas, or from
linearized quasi-normal modes on black brane backgrounds, which in the hydrodynamic limit correspond to shear and sound modes of the dual field theory. 
The standard holographic dictionary instructs us to extract correlators from the \emph{boundary} behavior of fluctuating bulk fields, appropriately supplemented with boundary conditions at the horizon. 
However,  hydrodynamics is an effective description of the system at long wavelengths and small frequencies, and thus one would expect it to be 
encoded in properties of the geometry and its fluctuations in the IR, i.e. near the horizon.
Thus, a natural question is to what extent the horizon of a black brane can \emph{fully} capture the hydrodynamic behavior of a strongly coupled plasma, and its transport properties.
Indeed, the diffusive modes can be understood \cite{Kovtun:2003wp,Starinets:2008fb} as fluctuations of the black brane horizon using the membrane 
paradigm \cite{Damour:1979wya,Thorne:1986iy}, which identifies the horizon with a fictitious fluid. This approach was made more precise in \cite{Iqbal:2008by} and led to 
various formulations for extracting transport coefficients entirely from the horizon geometry.
However, these methods have been limited to special cases.

In this paper we revisit some of these questions, and set up a universal -- and efficient -- framework for extracting the 
shear and bulk viscosities of strongly coupled
gauge theories with holographic duals
involving higher derivative corrections. 
A crucial step in our analysis is the realization that the terms needed to compute 
both $\eta$ and $\zeta$ can be extracted from  \emph{radially conserved currents}, even in the presence of higher derivatives. 
In turn, this implies that they can be evaluated 
at the black brane horizon. As we will see, one clear advantage of our framework is that 
it avoids having to compute dispersion relations.
In our analysis we will consider
both Weyl$^4$ corrections and Riemann$^2$ corrections, encoding, respectively, finite 't Hooft coupling and finite N effects.
Moreover, our gravitational dual has an arbitrary number of scalars,
with an arbitrary interaction potential.
Having such a framework is especially valuable for theories with higher derivatives, where the computations are intrinsically more cumbersome and a number of subtleties arise, related to the presence of, for instance, additional boundary terms and counterterms.

In closing we should mention that the recent paper \cite{Demircik:2023lsn} has also examined the connection between horizon and boundary data and has put forth an efficient method for computing 
transport coefficients directly from the horizon\footnote{See also \cite{Donos:2022uea,Davison:2022vqh,Donos:2022www} for related earlier treatment.}. However, their analysis is restricted to two-derivative theories, and to matter sectors involving, in addition to Einstein gravity and a 
$U(1)$ gauge field, only one scalar field. In our analysis, on the other hand, we have included an arbitrary number of scalars and allowed for curvature corrections to the leading gravitational action.

\subsection{Summary of Results}

We will work with a five-dimensional theory of gravity in AdS coupled to an arbitrary number of scalars, described by 
\begin{equation}
\begin{split}
S_5&=\frac{1}{16\pi G_N}\int_{\calm_5}d^5x \sqrt{-g}\ L_5
\\
&\equiv \frac{1}{16\pi G_N}\int_{\calm_5}d^5x \sqrt{-g}\ \biggl[
R+12-\frac{1}{2}\sum_i\left(\del \phi_i\right)^2-V\{\phi_i\}+\beta\cdot \dd\call
\biggr]\,,
\end{split}
\eqlabel{2der}
\end{equation}
where $\dd\call$ denote terms involving higher derivative corrections to Einstein gravity.
In particular, in this paper we consider two classes of models, to leading
order\footnote{Throughout the paper we keep the subscripts $ _2$ or $ _4$ in
reference to models \eqref{dl2} and \eqref{dl4}. The parameter $\beta$ is assumed to be perturbatively small. The asymptotic AdS radius is set $L=1$, in the absence of the higher derivative corrections.}
in $\beta$:
\begin{itemize}
\item
four-derivative curvature corrections described by: 
\begin{equation}
\dd\call_2\equiv \alpha_1\ R^2+\alpha_2\ R_{\mu\nu} R^{\mu\nu}
+\alpha_3 R_{\mu\nu\rho\lambda} R^{\mu\nu\rho\lambda}\,;
\eqlabel{dl2}
\end{equation}
\item
eight-derivative curvature corrections described by: 
\begin{equation}
\dd\call_4\equiv  
C^{hmnk} C_{pmnq} C_h\ ^{rsp}C^q\ _{rsk}+\frac 12 C^{hkmn}C_{pqmn}
C_h\ ^{rsp}C^q\ _{rsk}\,,
\eqlabel{dl4}
\end{equation}
where $C$ is the Weyl tensor.
\end{itemize}

For the shear viscosity to the entropy ratio we find, respectively:
\begin{itemize}
\item
\begin{equation}
\frac{\eta}{s}\bigg|_{\dd\call_2}= \frac{1}{4\pi}\biggl(1+\beta\cdot
\frac23\alpha_3 \left(V-12\right)\biggr)\,;
\eqlabel{etas2}
\end{equation}
\item
\begin{equation}
\frac{\eta}{s}\bigg|_{\dd\call_4}= \frac{1}{4\pi}\biggl(1-\beta\cdot
\frac{1}{72}(V-12)\biggl[
3 \sum_i (\del_i V)^2 +5(V-12)^2
\biggr]\biggr)\,,
\eqlabel{etas4}
\end{equation}
where $\del_i V\equiv \frac{\del V}{\del\phi_i}$.
\end{itemize}
All the quantifies in \eqref{etas2} and \eqref{etas4}
are to be evaluated at the horizon of the
dual black brane solution. Since the $\calo(\beta^0)$ results
are universal \cite{Buchel:2003tz}, it is sufficient
to evaluate the scalar potential and its derivatives to leading $\calo(\beta^0)$ order only.

For the bulk viscosity to the entropy ratio we find, respectively:
\begin{itemize}
\item
\begin{equation}
\begin{split}
9\pi \frac{\zeta}{s}\bigg|_{\dd\call_2}=\biggl(1&-\frac23 (V-12) (5 \alpha_1+\alpha_2-\alpha_3) \beta\biggr)
\sum_{i} z_{i,0}^2 \\&+\beta\cdot \frac{4(5 \alpha_1+\alpha_2-\alpha_3)} {3(V-12)}\ \sum_i (z_{i,0}\cdot\del_i V)^2\,;
\end{split}\eqlabel{bulks2}
\end{equation}
\item
\begin{equation}
\begin{split}
9\pi \frac{\zeta}{s}\bigg|_{\dd\call_4}= \biggl(1+\frac{5}{144} \beta (V-12)^3\biggr) \sum_{i} z_{i,0}^2
-\beta\cdot \frac{5}{24}(V-12)\ \sum_i (z_{i,0}\cdot\del_i V)^2\,.
\end{split}
\eqlabel{bulks4}
\end{equation}
\end{itemize}
Once again, all the quantifies in \eqref{bulks2} and \eqref{bulks4}
are to be evaluated at the horizon of the
dual black brane solution. Here $z_{i,0}$ are the values of the gauge invariant
scalar fluctuations, at zero frequency, evaluated at the black brane horizon,
see section \ref{zetaeta} and in particular \eqref{hydroz}. While the scalar potential
and its derivatives  can be evaluated to the leading $\calo(\beta^0)$
order of the background black brane solution, the horizon values of the scalars
$z_{i,0}$ must be evaluated including $\calo(\beta)$ corrections.

The rest of the paper is organized as follows.
In section \ref{checks} we present the
analysis for some specific models, and provide extensive checks on the
general formalism. We conclude in section \ref{conclude}
and highlight future directions. 
The formal proofs of the main results --- the final expressions \eqref{etas2},
\eqref{etas4} for the shear viscosity, and \eqref{bulks2}, \eqref{bulks4}
for the bulk viscosity are delegated to appendix \ref{proof}.
We discuss the background black 
brane geometry in \ref{background}, $\frac \eta s$ is computed
in section \ref{etas}, and  $\frac \zeta s$ is computed
in section \ref{zetaeta}.

\section{Applications}\label{checks}

In this section we use simple toy models to validate the general formulas
reported in \eqref{etas2}-\eqref{bulks4}.
First and foremost, note that if the boundary gauge theory is a CFT with
\begin{equation}
V\equiv 0\,,
\eqlabel{cftv}
\end{equation}
we find\footnote{The bulk viscosity of a CFT plasma vanishes;
see \cite{Benincasa:2005qc} for the original analysis of $\caln=4$ SYM.} from \eqref{etas2} and \eqref{etas4} 
\begin{equation}
\frac{\eta}{s}\bigg|_{\dd\call_2}^{CFT}=\frac{1}{4\pi}\left(1-\beta\cdot 8\alpha_3\right)\,,\qquad
\frac{\eta}{s}\bigg|_{\dd\call_4}^{CFT}=\frac{1}{4\pi}\left(1+\beta\cdot 120\right)\,,
\eqlabel{etas4cft}
\end{equation}
reproducing \cite{Kats:2007mq} and \cite{Buchel:2004di,Buchel:2008ac,Buchel:2008sh} correspondingly. 

We discuss the following models:
\begin{itemize}
\item $(\cala_{2,\Delta})$: $\dd\call_2$ model with
\begin{equation}
\{\alpha_1\,,\, \alpha_2\,,\, \alpha_3\}=\{0\,,\, 0\,,\, 1\}\,,
\eqlabel{asa2}
\end{equation}
with
\begin{equation}
V=\frac{m^2}{2}\ \phi^2\,,\qquad m^2\biggl(1+\frac 23\beta\biggr)=\Delta(\Delta-4)\,.
\eqlabel{va2}
\end{equation}
Note the $\calo(\beta)$ modification of the relation between the mass of the bulk scalar
and the dimension $\Delta$ of the dual boundary operator\footnote{See \cite{Buchel:2018ttd}
where the need for such a modification was first pointed out.}.
We consider $\Delta=\{2,3\}$.
The bulk viscosity in these models was not discussed in the literature before.

Models $\cala_{2,\Delta}$ are interesting in that the gravitational holographic bulk is higher derivative
{\it and} the black brane horizon Wald entropy differs from its Bekenstein entropy, see  \eqref{ss2}.
\item $(\calb_{2,\Delta})$: $\dd\call_2$ model with
\begin{equation}
\{\alpha_1\,,\, \alpha_2\,,\, \alpha_3\}=\{1\,,\, -4\,,\, 1\}\,,
\eqlabel{asb2}
\end{equation}
with
\begin{equation}
V=\frac{m^2}{2}\ \phi^2\,,\qquad m^2\biggl(1+2\beta\biggr)=\Delta(\Delta-4)\,.
\eqlabel{vb2}
\end{equation}
The $\calo(\beta)$ modification of the relation between the mass of the bulk scalar
and the dimension $\Delta$ of the dual boundary operator is precisely as reported in \cite{Buchel:2018ttd}.
We consider $\Delta=\{2,3\}$.
The bulk viscosity in these models was considered in \cite{Buchel:2018ttd}, but only to
leading order in the non-normalizable coefficient of the scalar $\phi$ (albeit to all
orders in $\beta$). Here we consider leading perturbative in $\beta$ corrections to 
transport in these models, but to all orders in the conformal symmetry breaking parameter,
\ie non-perturbatively in the non-normalizable coefficient of the bulk scalar $\phi$.

Models $\calb_{2,\Delta}$ are interesting in that the gravitational holographic bulk
represents a two-derivative model: the coefficients in \eqref{asb2} assemble
Riemann squared terms into the Gauss-Bonnet combination. Notice that for the black branes
in these models the Wald entropy is identical to their Bekenstein entropy, see
\eqref{ss2}.

\item $(\calc_{2,\Delta})$: $\dd\call_2$ model with
\begin{equation}
\{\alpha_1\,,\, \alpha_2\,,\, \alpha_3\}=\{0\,,\, 1\,,\, 1\}\,,
\eqlabel{asc2}
\end{equation}
with
\begin{equation}
V=\frac{m^2}{2}\ \phi^2\,,\qquad m^2\biggl(1+2\beta\biggr)=\Delta(\Delta-4)\,.
\eqlabel{vc2}
\end{equation}
The $\calo(\beta)$ modification of the relation between the mass of the bulk scalar
and the dimension $\Delta$ of the dual boundary operator is identical to
the one in models $(\calb_{2,\Delta})$.
We consider $\Delta=\{2,3\}$.
The bulk viscosity in these models was not discussed in the literature before.

Models $\calc_{2,\Delta}$ are interesting in that the gravitational holographic bulk
is higher-derivative, but the horizon physics is effectively two-derivative:
as in the case above, in these models there is no difference between the Wald and the
Bekenstein entropies of the dual black brane horizon.

\item $(\cald_{4,\Delta})$: $\dd\call_4$ model with
\begin{equation}
V=\frac{m^2}{2}\ \phi^2\,,\qquad m^2=\Delta(\Delta-4)\,.
\eqlabel{vd4}
\end{equation}
Notice that here the Weyl$ ^4$ higher derivative corrections to the gravitational
action \eqref{dl4} do not modify the bulk scalar mass/dimension of the dual operator
relation. We consider $\Delta=\{2,3\}$.
The bulk viscosity in these models was not discussed in the literature before.

Models $\cald_{4,\Delta}$ are interesting in that here
the higher derivative corrections are associated with 
finite 't Hooft coupling corrections of the UV fixed point CFT,
rather than with the 
difference between the central charges of the UV CFT, encoded by $\beta\cdot \alpha_3\equiv \frac{c-a}{8c}$,
as in models $\{\cala,\calb,\calc\}_{2,\Delta}$.

\end{itemize}

In the models just introduced we compute the shear and the bulk viscosities
using \eqref{etas2}-\eqref{bulks4},
and compare the results with direct computation of these quantities from the
dispersion relation of the shear and the sound modes,
\begin{equation}
\begin{split}
{\rm shear:}\qquad & \hw=-i\ \frac{2\pi\eta}{s}\ \hq^2+\calo(\hq^3)\,,
\\
{\rm sound:}\qquad & \hw=c_s\cdot \hq -i\ \frac{4\pi \eta}{3 s}\biggl(1+\frac{3\zeta}{4\eta}\biggr)\
\hq^2 +\calo(\hq^3)\,.
\end{split}
\eqlabel{disprel}
\end{equation}
These are the appropriate quasinormal modes of the dual black brane
\cite{Kovtun:2005ev}. The first non-conformal gauge theory computations of the bulk
viscosity from the dispersion relation were performed in \cite{Benincasa:2005iv}.
In genuinely higher-derivative holographic models the shear and the sound mode dispersion
relations where studied only in conformal $\caln=4$ SYM in \cite{Benincasa:2005qc}.
In this paper, we generalize (and combine) the computation methods of
\cite{Benincasa:2005iv} and \cite{Benincasa:2005qc}. Such analysis are much more involved 
and are extremely technical. We will not provide any details --- in fact our motivation
of developing the framework explained in sections \ref{etas} and \ref{zetaeta}
was precisely to avoid computation of the dispersion relations in the first place.
As we already emphasized, here we use such dispersion computations in models $\cala-\cald$
as a check on our general framework. 

Finally, we mention one additional test we performed. The speed of the sound waves $c_s$
in \eqref{disprel} is related to the equation of state $P=P(\cale)$ of the holographic
gauge theory plasma via
\begin{equation}
c_s^2=\frac{\del P}{\del \cale}\bigg|_{\lambda_\Delta={\rm const}}\,,
\eqlabel{cs2}
\end{equation}
where in computing derivatives of the pressure $P$ with respect to the energy density $\cale$
one has to keep the non-normalizable coefficient $\lambda_\Delta$ of the bulk scalar (\ie the coupling
constant of the dual operator $\calo_\Delta$ explicitly breaking the conformal invariance) constant.

As we explicitly show in this section, all the validations pass with excellence.

\subsection{Shear viscosity in models $\cala-\cald$}

Given \eqref{etas2}, we find that in all models $\cala-\calc$ the shear viscosity to the entropy
density is
\begin{equation}
\frac{\eta}{s}\bigg|_{\cala,\calb,\calc}=\frac{1}{4\pi}\left(1+\beta\cdot \dd_{\eta;\Delta}^{\cala,\calb,\calc}\right)\,,\qquad
\dd_{\eta;\Delta}^{\cala,\calb,\calc}=-8+\frac{\Delta(\Delta-4)}{3}\ (\phi_{0,0}^h)^2\,,
\eqlabel{etaac}
\end{equation}
while from \eqref{etas4} in models $\cald$ the shear viscosity to the entropy density is
\begin{equation}
\begin{split}
&\frac{\eta}{s}\bigg|_{\cald}=\frac{1}{4\pi}\left(1+\beta\cdot \dd_{\eta;\Delta}^\cald\right)
\,,\qquad \dd_{\eta;\Delta}^\cald=-\frac{1}{72}
\biggl(
\frac{\Delta (\Delta-4) (\phi_{0,0}^h)^2}{2}-12\biggr)\times\\
&\times 
\biggl(
5 \left(\frac12 \Delta (\Delta-4) (\phi_{0,0}^h)^2-12\right)^2+3 \Delta^2 (\Delta-4)^2 (\phi_{0,0}^h)^2
\biggr)\,,
\end{split}
\eqlabel{etad}
\end{equation}
where $\phi_{0,0}^h$ is the leading $\calo(\beta^0)$ order horizon value of the bulk scalar.
It is computed numerically solving the leading order $\calo(\beta^0)$ background equations of motion
\eqref{eq1}-\eqref{eq4}, using the metric parameterization \eqref{fixdiff}, subject to the boundary
conditions:
\begin{equation}
\begin{split}
&r\to 0:\qquad  f\sim 1+\cdots\,,\ \  g\sim 1+\cdots\,,\ \ \phi\sim \begin{cases}
\lambda_3\cdot r+\cdots\,,\qquad &{\rm when}\ \Delta=3\,;\\
\lambda_2\cdot r^2\ln r+\cdots\,,\qquad &{\rm when}\ \Delta=2\,,
\end{cases}\\
&r\to r_h:\qquad  f\sim 0 +\cdots\,,\qquad g\sim g_{0,0}^h+\cdots\,,\qquad \phi\sim \phi_{0,0}^h\,.
\end{split}
\eqlabel{uvirlead}
\end{equation}
Without loss of generality we can fix $r_h=1$, provided we present all results as dimensionless quantities.
From \eqref{deft} we find
\begin{equation}
2\pi T =\sqrt{g_{0,0}^h}\biggl( 2 + \frac{\Delta(4-\Delta)}{12}\ (\phi_{0,0}^h)^2\biggr)\,.
\eqlabel{tdelta}
\end{equation}

\begin{figure}[ht]
\begin{center}
\psfrag{x}[c]{{$m_f/(2\pi T)$}}
\psfrag{y}[cb]{{$\dd_{\eta;3}^{\cala,\calb,\calc}$}}
\psfrag{z}[c]{{$m_b^2/(2\pi T)^2$}}
\psfrag{u}[ct]{{$\dd_{\eta;2}^{\cala,\calb,\calc}$}}
  \includegraphics[width=3.0in]{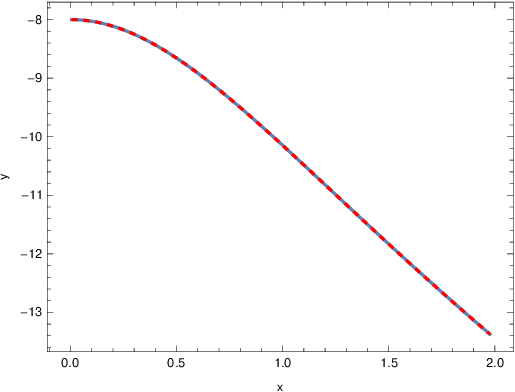}
  \includegraphics[width=3.0in]{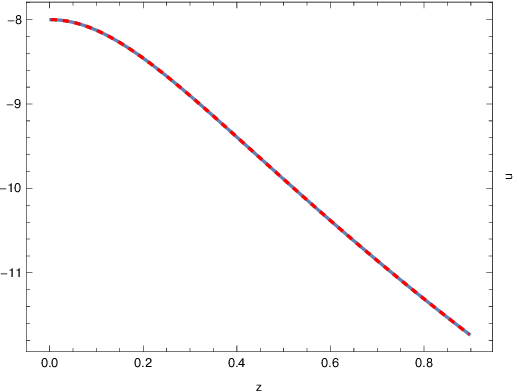}
\end{center}
 \caption{Corrections to the shear viscosity in models $\cala-\calc$ due to the UV fixed point
 central charges $c\ne a$ are universal, see \eqref{etaac}.
In the left panel we consider non-conformal gauge theories with a UV fixed point deformed
by a dimension $\Delta=3$ operator, $CFT\to CFT+m_f \calo_3$, while in the right panel
a UV fixed point is deformed by an operator of dimension $\Delta=2$, $CFT\to CFT+m_b^2 \calo_2$.
Solid curves represent the corrections to the shear viscosity extracted from the
shear channel quasinormal mode of the background black brane \eqref{disprel}, while the red dashed curves
are obtained applying \eqref{etaac}. 
}\label{figure1}
\end{figure}

\begin{figure}[ht]
\begin{center}
\psfrag{x}[c]{{$m_f/(2\pi T)$}}
\psfrag{y}[cb]{{$\dd_{\eta;3}^{\cald}$}}
\psfrag{z}[c]{{$m_b^2/(2\pi T)^2$}}
\psfrag{u}[ct]{{$\dd_{\eta;2}^{\cald}$}}
  \includegraphics[width=3.0in]{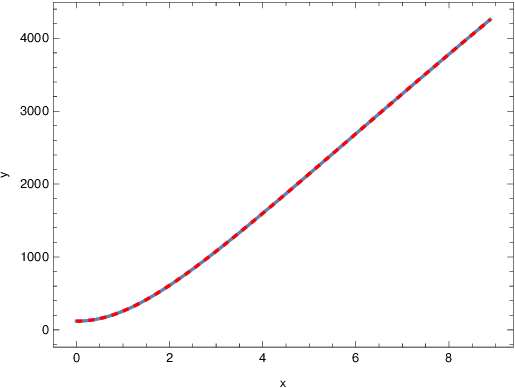}
  \includegraphics[width=3.0in]{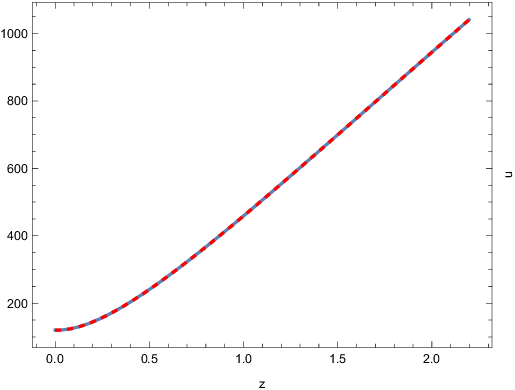}
\end{center}
 \caption{Corrections to the shear viscosity in models $\cald$ due to finite 't Hooft coupling corrections,
 see \eqref{etad}. In the left panel we consider non-conformal gauge theories with a UV fixed point deformed
by a dimension $\Delta=3$ operator, $CFT\to CFT+m_f \calo_3$, while in the right panel
a UV fixed point is deformed by an operator of dimension $\Delta=2$, $CFT\to CFT+m_b^2 \calo_2$.
Solid curves represent the corrections to the shear viscosity extracted from the
shear channel quasinormal mode of the background black brane \eqref{disprel}, while the red dashed curves
are obtained applying \eqref{etad}. 
}\label{figure2}
\end{figure}

Comparisons between the corrections to the shear viscosity (see \eqref{etaac} and \eqref{etad})
for the holographic models $\cala-\cald$ extracted from the dispersion relation of the shear modes using \eqref{disprel}
(the solid curves) and using \eqref{etaac} and \eqref{etad} (dashed red curves) are presented in
figs.~\ref{figure1}-\ref{figure2}. In the left panels we consider models with $\Delta=3$ with the
non-normalizable gravitational bulk scalar coefficient $\lambda_3$ identified as a fermionic mass term
$\lambda_3\equiv m_f$ of the boundary gauge theory. In the right panels we consider models with $\Delta=2$ with the
non-normalizable gravitational bulk scalar coefficient $\lambda_2$ identified as a bosonic mass term
$\lambda_2\equiv m_b^2$ of the boundary gauge theory\footnote{We adopted $m_f$ and $m_b^2$ labels
from \cite{Buchel:2007vy}.}. The difference between the solid and the dashed red curves is $\sim 10^{-8}\cdots 10^{-5}\ \%$
over the ranges of $\lambda_\Delta/T^{4-\Delta}$ reported. When $m_f=0$ and $m_b^2=0$ we recover the
conformal gauge theory results \eqref{etas4cft}:
\begin{equation}
\dd^{\cala,\calb,\calc}_{\eta;\Delta}\bigg|_{m_f=m_b^2=0}=-8\,,\qquad \dd^{\cald}_{\eta;\Delta}\bigg|_{m_f=m_b^2=0}=120\,.
\eqlabel{reccft}
\end{equation}

\subsection{Bulk viscosity in models $\cala-\cald$}

Unlike the shear viscosity, the bulk viscosity in models $\cala$ and $\calb,\calc$ differs: using \eqref{bulks2} we find
\begin{equation}
\begin{split}
&\frac{\zeta}{\eta}\bigg|_{\cala}=\frac49\ z_{0,0}^2+\beta\cdot \dd_{\zeta;\Delta}^{\cala}\,,\\
&\dd_{\zeta;\Delta}^{\cala}=\frac89 z_{0,0} z_{0,1}+\frac{8}{27} \biggl(\frac12 \Delta (\Delta-4) (\phi_{0,0}^h)^2-12\biggr)
z_{0,0}^2-\frac{16}{27}\ \frac{  \Delta^2 (\Delta-4)^2 (\phi_{0,0}^h)^2}{\frac12 \Delta (\Delta-4) (\phi_{0,0}^h)^2-12}\
z_{0,0}^2
\\&-\frac49 z_{0,0}^2\ \dd_{\eta;\Delta}^{\cala,\calb,\calc}\,,
\end{split}
\eqlabel{bulka}
\end{equation}
\begin{equation}
\begin{split}
&\frac{\zeta}{\eta}\bigg|_{\calb,\calc}=\frac49\ z_{0,0}^2+\beta\cdot \dd_{\zeta;\Delta}^{\calb,\calc}\,,\qquad
\dd_{\zeta;\Delta}^{\calb,\calc}=\frac89 z_{0,0} z_{0,1}-\frac49 z_{0,0}^2\ \dd_{\eta;\Delta}^{\cala,\calb,\calc}\,,
\end{split}
\eqlabel{bulkbc}
\end{equation}
and
\begin{equation}
\begin{split}
&\frac{\zeta}{\eta}\bigg|_{\cald}=\frac49\ z_{0,0}^2+\beta\cdot \dd_{\zeta;\Delta}^\cald\,,\\
&\dd_{\zeta;\Delta}^{\cald}=\frac49 z_{0,0}^2\
\biggl[ \frac{5}{144} \left(\frac12 \Delta (\Delta-4) (\phi_{0,0}^h)^2-12\right)^3
\\&-\frac{5}{24} \left(\frac 12 \Delta (\Delta-4) (\phi_{0,0}^h)^2-12\right) \Delta^2 (\Delta-4)^2
(\phi_{0,0}^h)^2\biggr]
+\frac89 z_{0,0} z_{0,1}-\frac49 z_{0,0}^2 \dd_{\eta;\Delta}^{\cald}\,,
\end{split}
\eqlabel{bulkd}
\end{equation}
where $\phi_{0,0}^h$ is the leading $\calo(\beta^0)$ order horizon value of the bulk scalar.
In \eqref{bulka}, \eqref{bulkbc}  and \eqref{bulkd} we denoted
\begin{equation}
z_{i,0}\equiv z_{0,0}+\beta\cdot z_{0,1}\,,
\eqlabel{z00z01}
\end{equation}
since our toy models have a single bulk scalar. 
To proceed further we need to evaluate the gauge invariant bulk scalar fluctuations $z_0$ at the horizon
to order $\calo(\beta)$ as emphasized in \eqref{z00z01}.

We present details for the model $\cala_{2,\Delta=3}$, and only the final results for the other models.  

\subsubsection{Model $\cala_{2,\Delta=3}$}\label{moda}

Since we will need the gauge invariant bulk scalar fluctuations $z_0$ at the horizon
to order $\calo(\beta)$, we need the background geometry to order $\calo(\beta)$.
It is convenient to use the metric warp-factor parameterization as in
\eqref{fixdiff}. Explicitly,
\begin{equation}
c_1=\frac{f^{1/2}\sqrt{g+\beta\cdot g_{1,1}}}{r}\,,\ c_2=\frac1r\,,\ c_3=\frac{1-\frac 13\beta}{r f^{1/2}(1
+\beta\cdot g_{2,1})}\,,\ \phi=\phi_0+\beta\cdot \phi_1 \,.
\eqlabel{c1c3def}
\end{equation}
From \eqref{eq1}-\eqref{eq4} we find
\begin{equation}
\begin{split}
&0=f'-\frac r6 (\phi_0')^2 f+\frac{\phi_0^2}{2r}-\frac{4 f}{r}+\frac 4r\,,
\end{split}
\eqlabel{bac1}
\end{equation}
\begin{equation}
\begin{split}
&0=g'+ \frac g3 (\phi_0')^2 r\,,
\end{split}
\eqlabel{bac2}
\end{equation}
\begin{equation}
\begin{split}
&0=\phi_0''+\left(\frac 1r- \frac{\phi_0^2}{2r f}-\frac{4}{r f}\right) \phi_0'+\frac{3 \phi_0}{r^2 f}\,,
\end{split}
\eqlabel{bac3}
\end{equation}
at order $\calo(\beta^0)$, and
\begin{equation}
\begin{split}
&0=g_{2,1}'-\left(\frac{\phi_0^2}{2r f}+\frac{4}{r f}\right) g_{2,1}
-\frac r6 \phi_0' \phi_1'-\frac{fr^5}{72} (\phi_0')^6
-\frac{r^3}{24} (2 \phi_0^2+f+16) (\phi_0')^4
\\&+\frac49 \phi_0 r^2 (\phi_0')^3+\biggl(
\frac{11}{3} r f+2 r-\frac{8 r}{f}-\frac{r}{8f} \phi_0^4-\left(\frac{2r}{f}-\frac{r}{12}\right) \phi_0^2
\biggr) (\phi_0')^2
\\&+\biggl(
\frac{4\phi_0^3}{3f}+\frac{4(8-4 f) \phi_0}{3f}\biggr) \phi_0'-\frac{\phi_0^4}{24r f}
+\left(\frac1r-\frac{14}{3 r f}\right) \phi_0^2+\frac{\phi_0 \phi_1}{2r f}
-\frac{4 (f-1)^2}{r f}\,,
\end{split}
\eqlabel{bac4}
\end{equation}
\begin{equation}
\begin{split}
&0=g_{1,1}'+\frac r3 (\phi_0')^2  g_{1,1}+\frac {g r}{3} \phi_0'\phi_1'
-\frac{g (\phi_0^2+8)}{r f} g_{2,1}+\frac{g \phi_0}{r f} \phi_1
+\frac{f r^5 g}{108} (\phi_0')^6+\frac{r^3 g}{36} (2 \phi_0^2\\
&-11 f+16) (\phi_0')^4
-\frac29 g \phi_0 r^2 (\phi_0')^3+
+\frac{r g}{12f} \left(\phi_0^4-2 \phi_0^2 (3 f-8)+72 f^2-48 f+64\right) (\phi_0')^2
\\&-\frac{2g \phi_0 (\phi_0^2+8 f+8)}{3f} \phi_0'
-\frac{g}{12r f} \left(\phi_0^4-8 \phi_0^2 (3 f-2)+96 (f-1)^2\right)\,,
\end{split}
\eqlabel{bac5}
\end{equation}
\begin{equation}
\begin{split}
&0=\phi_1''+\biggl(\frac1r-\frac{\phi_0^2}{2r f}-\frac{4}{r f}\biggr) \phi_1'
-\frac{r \phi_0 \phi_0'-3}{r^2f} \phi_1-g_{2,1} \frac{6 \phi_0-r (\phi_0^2+8) \phi_0'}{r^2 f}
\\&+\frac{f r^5}{108} (\phi_0')^7+\frac{r^3}{36} (2 \phi_0^2+7 f+16) (\phi_0')^5
-\frac{r^2 \phi_0}{3} (\phi_0')^4+\frac{r }{12f}(\phi_0^4+2 \phi_0^2 (f+8)-80 f^2\\
&+64) (\phi_0')^3
-\frac{\phi_0 (\phi_0^2-8 f+8)}{f} (\phi_0')^2
+\frac{\phi_0'}{12rf} (\phi_0^4-8 \phi_0^2 (3 f-8)+96 (f-1)^2)\,,
\end{split}
\eqlabel{bac6}
\end{equation}
at order $\calo(\beta)$. The background equations \eqref{bac1}-\eqref{bac6} are solved
subject to the following asymptotics:
\nxt near the AdS boundary, \ie as $r\to 0$,
\begin{equation}
\begin{split}
&\phi_0=\lambda_3 r+r^3 \left(\phi_{0;3}+\frac16 \lambda_3^3 \ln r\right)+\calo(r^5\ln r)\,,\qquad
g=1-\frac16 r^2 \lambda_3^2
+\calo(r^4\ln r)\,,\\
&f=1+\frac16 r^2 \lambda_3^2
+r^4 \left(f_{4}+\frac{1}{12} \lambda_3^4 \ln r\right)+\calo(r^6 \ln^2 r)\,,\\
&\phi_1=r^3\left(\phi_{1;3}-\frac23\lambda_3^3 \ln r\right) +\calo(r^5\ln r)\,,\qquad
g_{1,1}=r^4\biggl(-\frac13\lambda_3^4-\frac23\lambda_3 \phi_{0;3}-\frac12\lambda_3 \phi_{1;3}
\\&+2g_{2,1;4}\biggr)+\calo(r^6)\,,\qquad g_{2,1}=-\frac 13 r^2\lambda_3^2
+r^4\biggl(g_{2,1;4}-\frac19 \lambda_3^4\ln r\biggr)+\calo(r^6\ln^2 r)\,,
\end{split}
\eqlabel{uva}
\end{equation}
where $\lambda_3\equiv m_f$ is the non-normalizable coefficient of the bulk
scalar, and the coefficients $\{\phi_{0;3}, f_{4}, \phi_{1;3}, g_{2,1;4}\}$ are related
to the thermal expectation values of various boundary gauge theory operators;
\nxt in the vicinity of the black brane horizon, \ie as $y\equiv (1-r)\to 0$,
\begin{equation}
\begin{split}
&\phi_0=\phi_{0,0}^h+\calo(y)\,,\qquad g=g_{0,0}^h+\calo(y)\,,\qquad f=\left(4+\frac 12 (\phi_{0,0}^h)^2\right) y +\calo(y^2)\,,\\
&\phi_1=\phi_{1,0}^h+\calo(y)\,,\qquad g_{1,1}=g_{1,1;0}^h+\calo(y)\,,\qquad
\\&g_{2,1}=-\frac{(\phi_{0,0}^h)^4+28(\phi_{0,0}^h)^2-12\phi_{0,0}^h\phi_{1,0}^h+96}{12
((\phi_{0,0}^h)^2+8)} +\calo(y)\,,
\end{split}
\eqlabel{ira}
\end{equation}
specified by the set of coefficients $\{\phi_{0,0}^h,g_{0,0}^h,\phi_{1,0}^h,g_{1,1;0}^h\}$.

Using \eqref{ira}, from \eqref{deft} we compute
\begin{equation}
\begin{split}
&{2\pi T}\equiv s_0+\beta\cdot s_1\,,\qquad s_0= \frac{(g_{0,0}^h)^{1/2}}{4}\biggl((\phi_{0,0}^h)^2+8\biggr)\,,\qquad 
\\&s_1=\frac{(g_{0,0}^h)^{-1/2}}{48}\biggl[
12 g_{0,0}^h \phi_{0,0}^h \phi_{1,0}^h
+6 g_{1,1;0}^h \biggl((\phi_{0,0}^h)^2+8\biggr)-g_{0,0}^h \biggl((\phi_{0,0}^h)^4+24 (\phi_{0,0}^h)^2+64\biggr)\biggr]\,.
\end{split}
\eqlabel{defmta}
\end{equation}
It is important that $s_1\ne 0$, since as we will show it affects the representation of $\dd_{\zeta;\Delta=3}^\cala$
in the plots.

We continue with the equations of motion for the gauge invariant fluctuations
$z_{0,0}$ (at leading order in $\beta$) and $z_{0,1}$ at order $\calo(\beta)$, see \eqref{zieoms},
\eqref{hydroz} and \eqref{z00z01},
\begin{equation}
\begin{split}
&0=z_{0,0}''-\frac{\phi_0^2-2 f+8}{2r f} z_{0,0}'+\frac{z_{0,0}}{6r^2 f}
\biggl((\phi_0')^2 r^2 (\phi_0^2+8)-12 r \phi_0 \phi_0'+18\biggr)\,,
\end{split}
\eqlabel{z00}
\end{equation}
\begin{equation}
\begin{split}
&0=z_{0,1}''-\frac{\phi_0^2-2 f+8}{2r f} z_{0,1}'+\frac{z_{0,1}}{6r^2 f}
\biggl((\phi_0')^2 r^2 (\phi_0^2+8)-12 r \phi_0 \phi_0'+18\biggr)
+\biggl[
\frac{7f r^5}{108} (\phi_0')^6
\\&+\frac{5r^3}{36} (4 \phi_0^2+7 f+32) (\phi_0')^4
-3 \phi_0 r^2 (\phi_0')^3+\frac {r}{4f} \biggl(
3 \phi_0^4-6 \phi_0^2 (f-8)\\&-16 (f+2) (5 f-6)\biggr) (\phi_0')^2
+\frac{1}{108r f} \biggl(-684 \phi_0^3 r+144 r \phi_0 (21 f-38)\biggr) \phi_0'
+\frac{1}{12rf} \biggl(
\phi_0^4\\&+4 \phi_0^2 (3 g_{2,1}-6 f+40)-12 \phi_1 \phi_0+96 g_{2,1}+96 (f-1)^2
\biggr)\biggr] z_{0,0}'
+\biggl[
\frac{f r^6}{324} (\phi_0')^8+\frac{r^4}{108} (4 \phi_0^2\\
&+7 f+32) (\phi_0')^6
-\frac{\phi_0 r^3}{9} (\phi_0')^5
+\frac{r^2}{36f} \biggl(
\phi_0^4-22 \phi_0^2 f+16 \phi_0^2-80 f^2-188 f+64\biggr) (\phi_0')^4
\\&+\frac{\phi_0 r}{2f} (\phi_0^2+8 f+8) (\phi_0')^3
-\frac{1}{36f} \biggl(
7 \phi_0^4+4 \phi_0^2 (3 g_{2,1}-6 f+88)-12 \phi_1 \phi_0-96 f^2\\&
+96 g_{2,1}-480 f+1152
\biggr) (\phi_0')^2+\frac{r (\phi_0^2+8) \phi_0'-6 \phi_0}{3r f} \phi_1'
+\frac{1}{6f^2 r^2} \biggl(
\phi_0^5 r+4 r (f+4) \phi_0^3\\&+8 r (3 g_{2,1} f-6 f^2+16 f+8) \phi_0-12 f \phi_1 r\biggr) \phi_0'
-\frac{\phi_0^4+6 g_{2,1} f-4 \phi_0^2 (f-2)}{f^2 r^2}\biggr] z_{0,0}\,.
\end{split}
\eqlabel{z01}
\end{equation}
The fluctuations $z_{0,0}$ and $z_{0,1}$ are then solved subject to the following asymptotics: 
\nxt near the AdS boundary, \ie as $r\to 0$,
\begin{equation}
z_{0,0}=\frac12\lambda_3 r+r^3\biggl(z_{0,0;3}+\frac14 \lambda_3^3\ln r\biggr)+\calo(r^5\ln r)\,,
\  z_{0,1}=r^3\biggl(z_{0,1;3}-\lambda_3^3\ln r\biggr)+\calo(r^5\ln r)\,,
\eqlabel{zuva}
\end{equation}
where we note that the non-normalizable coefficients of $\{z_{0,0},z_{0,1}\}$, \ie
$\{\frac{1}{2}\lambda_3,0\}$
are precisely as required by \eqref{zibads};
\nxt in the vicinity of the black brane horizon, \ie as $y\equiv (1-r)\to 0$,
\begin{equation}
z_{0,0}=z_{0,0;0}^h+\calo(y)\,,\qquad z_{0,1}=z_{0,1;0}^h+\calo(y)\,.
\eqlabel{zira}
\end{equation}

Even though we will not present the equations for $z_{i,1}\equiv z_1\equiv z_{1,0}+\beta\cdot z_{1,1}$,
we solved them as well, to validate the conservation of the imaginary part of the
current \eqref{jbulk}
along the radial flow. Specifically, see \eqref{jwbulk},
\begin{equation}
\begin{split}
\lim_{r\to 0} F\equiv F^{(b)} =\frac 12 \lambda_3 z_{1,0;3}\,,\quad \lim_{r\to r_h=1} F\equiv F^{(h)}
=\frac{(g_{0,0}^h)^{1/2}}{8}\biggl((\phi_{0,0}^h)^2+8\biggr) (z_{0,0;0}^h)^2\,,
\end{split}
\eqlabel{fmoda}
\end{equation}
where $z_{1,0;3}$ is the normalizable coefficient of $z_{1,0}$ (similar to $z_{0,0;3}$ of $z_{0,0}$ in
\eqref{zuva}), and
\begin{equation}
\begin{split}
&\lim_{r\to 0} \dd F\equiv \dd F^{(b)} =\left(\frac 16 z_{1,0;3}+\frac 12 z_{1,1;3}\right)\lambda_3
\,,\quad \lim_{r\to r_h=1} \dd F\equiv \dd F^{(h)}
=(z_{0,0;0}^h)^2 \biggl(
\frac{(g_{0,0}^h)^{1/2}}{96} \biggl(\\
&-(\phi_{0,0}^h)^4+72 (\phi_{0,0}^h)^2-64\biggr)
+\frac{g_{1,1;0}^h}{16(g_{0,0}^h)^{1/2}} \biggl((\phi_{0,0}^h)^2+8\biggr)
+\frac18 (g_{0,0}^h)^{1/2} (\phi_{0,0}^h) \phi_{1,0}^h
\biggr)
\\&+\frac14 z_{0,0;0}^h (g_{0,0}^h)^{1/2} z_{0,1;0}^h \biggl((\phi_{0,0}^h)^2+8\biggr)\,,
\end{split}
\eqlabel{dfmoda}
\end{equation}
where $z_{1,1;3}$ is the normalizable coefficient of $z_{1,1}$ (similar to $z_{0,0;3}$ of $z_{0,0}$ in
\eqref{zuva}). Conservation of the imaginary part of current \eqref{jbulk} in particular requires that
\begin{equation}
\frac{F^{(b)}}{F^{(h)}}-1=0\,,\qquad \frac{\dd F^{(b)}}{\dd F^{(h)}}-1=0\,,
\eqlabel{curconst}
\end{equation}
and provides a stringent test on our numerics.

\begin{figure}[ht]
\begin{center}
\psfrag{x}[c]{{$m_f/(2\pi T)$}}
\psfrag{y}[cb]{{$\frac{\zeta}{\eta}|_0$}}
\psfrag{u}[ct]{{$\dd_{\zeta;3}^{\cala}$}}
  \includegraphics[width=3.0in]{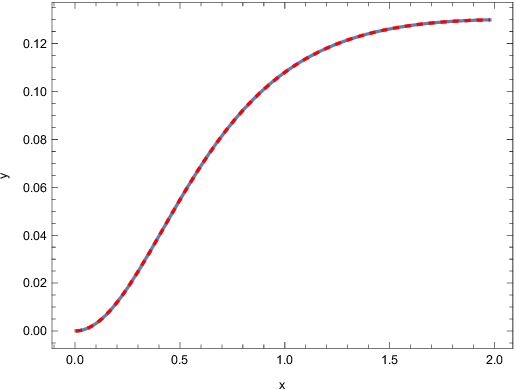}
  \includegraphics[width=3.0in]{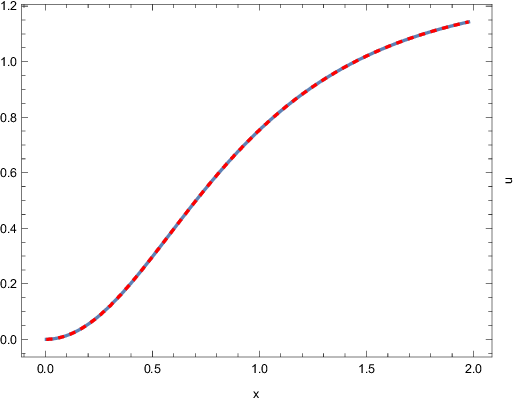}
\end{center}
\caption{Bulk viscosity in model $(\cala_{2,\Delta=3})$, \eqref{asa2}.   
The solid curves represent the leading $\calo(\beta^0)$ order  bulk viscosity (the left panel)
and its $\calo(\beta)$ correction (the right panel) extracted from the
sound wave channel quasinormal mode of the background black brane \eqref{disprel}. The red dashed curves
are obtained from \eqref{bulka}. 
}\label{figure3}
\end{figure}

\begin{figure}[ht]
\begin{center}
\psfrag{x}[c]{{$\lambda_3\equiv m_f$}}
\psfrag{y}[cb]{{${F^{(b)}}/{F^{(h)}}-1$}}
\psfrag{u}[ct]{{${\dd F^{(b)}}/{\dd F^{(h)}}-1$}}
  \includegraphics[width=3.0in]{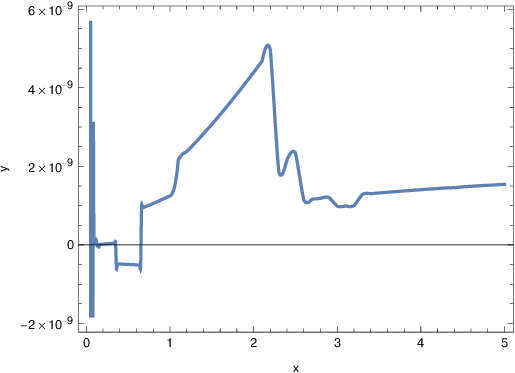}
  \includegraphics[width=3.0in]{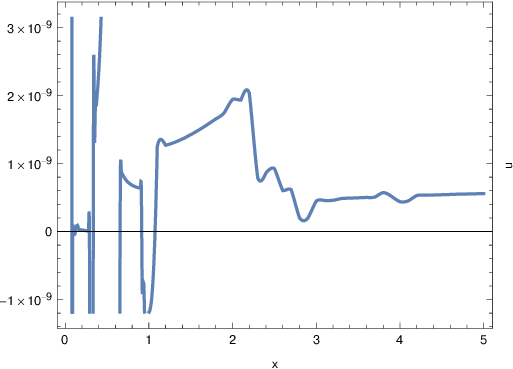}
\end{center}
 \caption{Numerical test of the conservation of the imaginary part of 
 current \eqref{jbulk} in model $(\cala_{2,\Delta=3})$
to leading order $\calo(\beta^0)$ (the left panel), and to subleading order $\calo(\beta)$ 
(the right panel). See \eqref{curconst} for more details. 
}\label{figure4}
\end{figure}

In practice we solve numerically \eqref{bac1}-\eqref{bac6}, \eqref{z00}-\eqref{z01}, along with
the equations for $z_{1,0}$ and $z_{1,1}$, parameterized by $\lambda_3\equiv m_f$. We use
\eqref{bulka} to extract the bulk viscosity, and compare the results with the quasinormal
modes computations\footnote{As we already mentioned, these computations are too technical
to report in details here.} \eqref{disprel}. 
It is important to present the physical results
as dimensionless quantities, as we fixed the overall scale on the gravitational side of the
computations setting $r_h=1$. From \eqref{defmta}, the dimensionless quantity $m_f/T$ is $\calo(\beta)$
corrected,
\begin{equation}
\frac{m_f}{2\pi T} \equiv x(\lambda_3)+\beta\cdot \dd x(\lambda_3)\,.
\eqlabel{corre}
\end{equation}
Assume that we have a dimensionless quantity $\calk$ that is $\calo(\beta)$ corrected, and that 
is extracted from the numerics as a functions of $\lambda_3$, but we need to present it as a function
of $m_f/(2\pi T)$. Then,
\begin{equation}
\calk=\calk_0(x+\beta\cdot \dd x)+\beta\cdot \calk_1(x+\beta\cdot \dd x) = \calk_0(x)+\beta\cdot
\biggl(\dd x\cdot \frac{d \calk_0(x)}{dx}+\calk_1(x)\biggr)\,,
\eqlabel{cork}
\end{equation}
\ie the $\calo(\beta)$ correction of the quantity $\calk$ receives an extra derivative in the $\calk_0$
term. 
This was not an issue in our discussion of the shear viscosity to the entropy density ratio
in section \ref{etas}, since there the appropriate quantity $\calk_0\equiv \frac{1}{4\pi}$ is a
constant. 

In fig.~\ref{figure3} we compare the leading $\calo(\beta^0)$ (the left panel) and the subleading
$\calo(\beta)$ correction (the right panel) of the ratio of the bulk viscosity to shear viscosity
using the formalism of section \ref{zetaeta} (the red dashed curves), and the
same quantities obtained from the computation of the sound channel quasinormal mode
of the background black brane \eqref{disprel}. The difference between the solid and the dashed red curves is
$\sim 10^{-6}\cdots 10^{-4}\ \%$.

In fig.~\ref{figure4} we numerically validate the conservation of the imaginary part of 
current \eqref{jbulk}.

\subsubsection{Models $\cala-\cald$}\label{modad}

\begin{figure}[ht]
\begin{center}
\psfrag{x}[c]{{$m_f/(2\pi T)$}}
\psfrag{t}[c]{{$m_b^2/(2\pi T)^2$}}
\psfrag{y}[cb]{{$\frac{\zeta}{\eta}|_0$}}
\psfrag{u}[ct]{{$\frac{\zeta}{\eta}|_0$}}
  \includegraphics[width=3.0in]{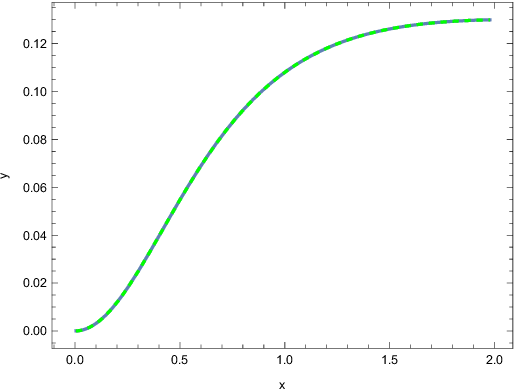}
  \includegraphics[width=3.0in]{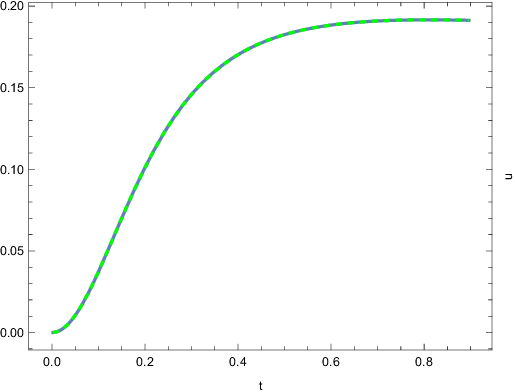}
\end{center}
\caption{To order  $\calo(\beta^0)$ there is a perfect agreement between the ratio of  bulk viscosity to the
shear viscosity evaluated using our novel formula \eqref{our}, shown in the solid curves,
and the Eling-Oz expression \eqref{zeeo}, shown in the dashed green curves.
}\label{figure5}
\end{figure}

\begin{figure}[ht]
\begin{center}
\psfrag{x}[c]{{$m_f/(2\pi T)$}}
\psfrag{y}[cb]{{$\dd_{\zeta,3}^\cala$}}
\psfrag{z}[ct]{{$\dd_{\zeta,3}^\calb$}}
  \includegraphics[width=3.0in]{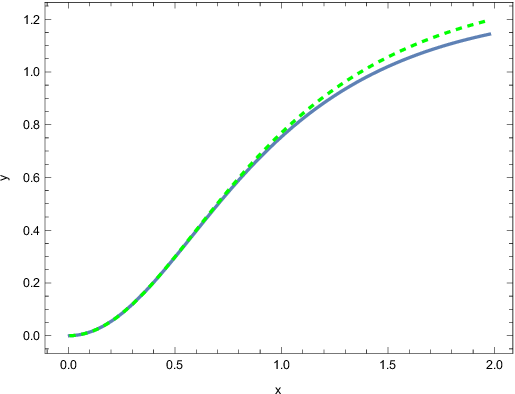}
\includegraphics[width=3.0in]{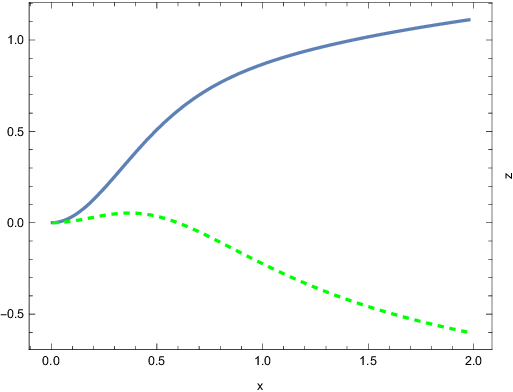}
\end{center}
\caption{At order  $\calo(\beta)$ there is a  disagreement between the ratio of bulk viscosity to the
shear viscosity evaluated using  \eqref{bulka} (the left panel) and \eqref{bulkbc} (the right panel),
shown in the solid curves,
and the extension of the Eling-Oz formula \eqref{zeeo}  to order $\calo(\beta)$, shown in the dashed green curves.
}\label{figure6}
\end{figure}

\begin{figure}[ht]
\begin{center}
\psfrag{x}[c]{{$m_f/(2\pi T)$}}
\psfrag{y}[ct]{{$\dd\ [{\zeta}/{s}]$}}
  \includegraphics[width=4.0in]{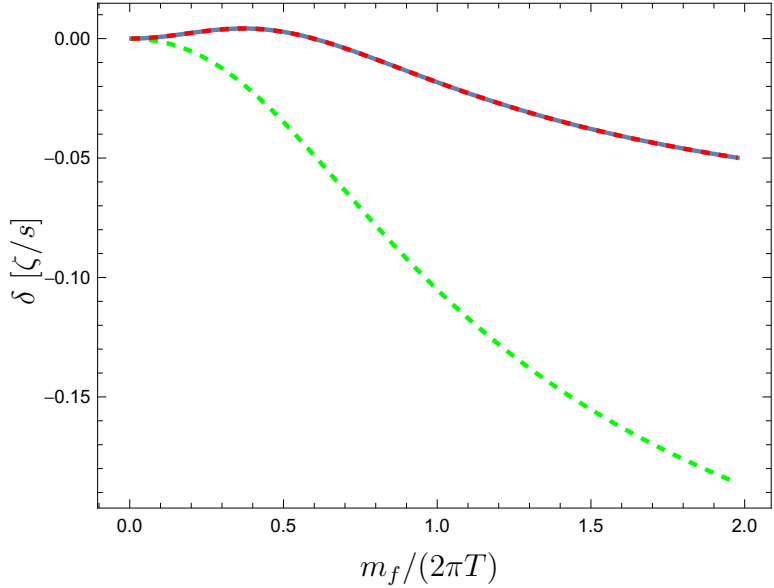}
\end{center}
\caption{We compare $\calo(\beta)$ correction to $\frac \zeta s$ in the
holographic model $\calb_{2,3}$: solid curve is obtained from the quasinormal mode \eqref{disprel}
analysis, the red dashed curve is obtained from \eqref{our2}, and the green dashed curve is the
application of the EO formula \eqref{zetaseo}.
}\label{fig6a}
\end{figure}

\begin{figure}[ht]
\begin{center}
\psfrag{x}[c]{{$m_f/(2\pi T)$}}
\psfrag{t}[c]{{$m_b^2/(2\pi T)^2$}}
\psfrag{y}[cb]{{$\dd_{\zeta,3}^{\cala,\calb,\calc}$}}
\psfrag{u}[ct]{{$\dd_{\zeta,2}^{\cala,\calb,\calc}$}}
  \includegraphics[width=3.0in]{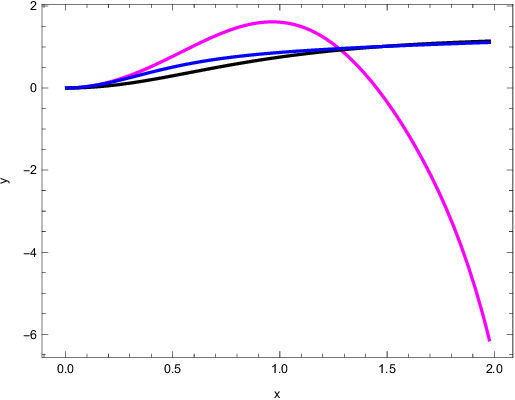}
\includegraphics[width=3.0in]{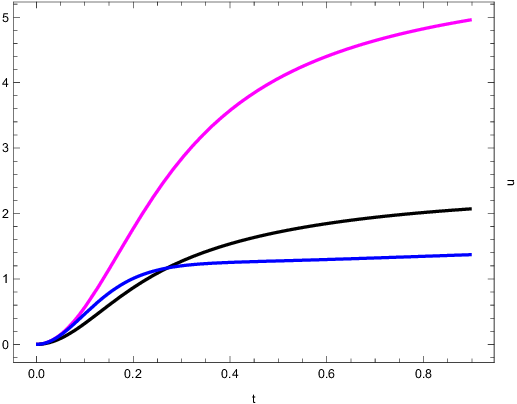}
\end{center}
\caption{Although models $\cala-\calc$ have the same value of the higher derivative coupling
$\alpha_3=1$, and the coupling constants $\alpha_1$ and $\alpha_2$ can be removed by a metric redefinition,
such a redefinition modifies the scalar sector of the model resulting in distinct
bulk viscosity corrections.
 Black curves represent $\cala_{2,\Delta}$ models, blue curves
represent $\calb_{2,\Delta}$ models, and magenta curves represent $\calc_{2,\Delta}$ models. 
}\label{figure7}
\end{figure}

\begin{figure}[ht]
\begin{center}
\psfrag{x}[c]{{$m_f/(2\pi T)$}}
\psfrag{t}[c]{{$m_b^2/(2\pi T)^2$}}
\psfrag{y}[cb]{{$\dd_{\zeta,3}^{\cald}$}}
\psfrag{u}[ct]{{$\dd_{\zeta,2}^{\cald}$}}
  \includegraphics[width=3.0in]{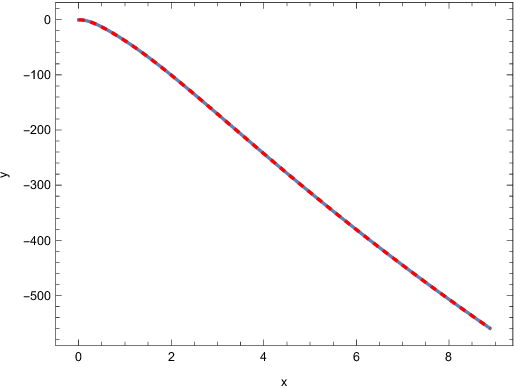}
\includegraphics[width=3.0in]{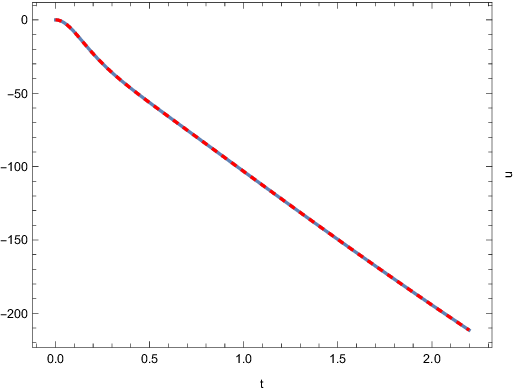}
\end{center}
\caption{Bulk viscosity corrections in the $\cald_{4,3}$ model (the left panel) and in the
$\cald_{4,2}$ model (the right panel). Solid curves represent corrections extracted from the
sound wave
channel quasinormal mode of the background black brane \eqref{disprel}. The red curves are
obtained from \eqref{bulkd}.
}\label{figure8}
\end{figure}

\begin{figure}[ht]
\begin{center}
\psfrag{x}[c]{{$m_f/(2\pi T)$}}
\psfrag{t}[c]{{$m_f/(2\pi T)$}}
\psfrag{y}[cb]{{$(\beta_{1,0})^2$}}
\psfrag{u}[ct]{{$\beta_{1,1}$}}
  \includegraphics[width=3.0in]{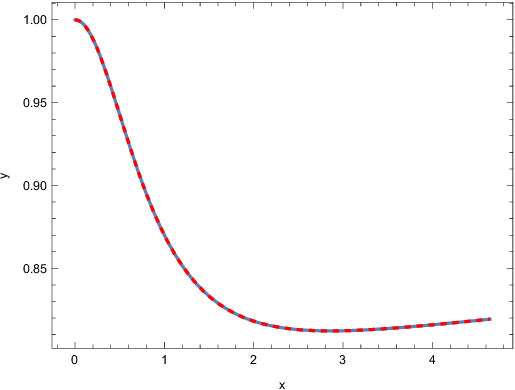}
\includegraphics[width=3.0in]{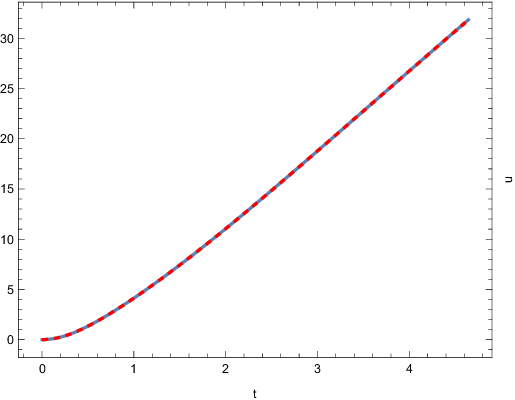}
\end{center}
\caption{We parameterize the speed of the sound waves in the gauge theory plasma as in \eqref{soundd3}.
The solid curves represent $(\beta_{1,0})^2$ and $\beta_{1,1}$ in the holographic model $\cald_{4,3}$
obtained from the sound wave channel quasinormal mode \eqref{disprel}; the dashed red curves
represent the same data obtained applying the equation of state \eqref{cs2}.
}\label{figure9}
\end{figure}

Analysis of the remaining models proceeds as detailed for the model $(\cala_{2,\Delta=3})$
in section \ref{moda}. As for model $(\cala_{2,\Delta=3})$, there is an excellent agreement
between the general computation framework of section \ref{zetaeta} and the alternative
extraction of the bulk viscosity from the quasinormal modes \eqref{disprel}.

A simple formula to compute the bulk viscosity in two-derivative holographic models
was proposed in \cite{Eling:2011ms} (EO):
\begin{equation}
\frac{\zeta}{\eta}=c_s^4 \sum_i \left(\frac{\del \phi_{i}}{\del \ln T}\right)^2\bigg|_{\lambda_{\Delta}={\rm const}} \,,
\eqlabel{zeeo}
\end{equation}
where $c_s$ is the speed of the sound waves in the holographic plasma \eqref{cs2},  and the
bulk scalar derivatives are evaluated at the black brane horizon, keeping the non-normalizable
coefficients of these scalars --- the mass terms of the boundary gauge theory --- fixed
\cite{Buchel:2011yv,Buchel:2011wx}. The EO was extensively tested in many models, and
it is verified to leading order $\calo(\beta^0)$ in the models discussed here. Specifically,
in fig.~\ref{figure5} we compare $\frac{\zeta}{\eta}\bigg|_0$
for $\cala-\cald$ models with $\Delta=3$ and $\Delta=2$  (the green gashed curves)
with the predictions \eqref{bulks2} of the framework discussed in section \ref{zetaeta}
(the solid curves):
\begin{equation}
\frac{\zeta}{s}\bigg|_0=\frac{1}{9\pi} \sum_{i} z_{i,0}^2\qquad \Longrightarrow\qquad 
\frac{\zeta}{\eta}\bigg|_0=\frac{4}{9} \sum_{i} z_{i,0}^2\,.
\eqlabel{our}
\end{equation}

We stress that \eqref{our} is a simple, novel expression for the bulk viscosity in two-derivative
holographic models: unlike  \cite{Gubser:2008sz}, there is no restriction to a single bulk scalar,
and there is no need to compute scalar derivatives at the horizon --- $z_{i,0}$ are values
of the gauge invariant scalar fluctuations\footnote{From the
quasinormal mode perspective, the $z_i$'s are spatially $SO(3)$ invariant background scalar fluctuations
of the sound channel which decouple from the metric fluctuations as $\kk\to 0$.}
at zero frequency evaluated at the horizon. Furthermore, the first expression in \eqref{our}
is true even in higher-derivative holographic models, provided the dual gravitational physics is effectively
two-derivative
(as in the Gauss-Bonnet models $\calb_{(2,\Delta)}$), or, at least, effectively two-derivative
at the horizon\footnote{Recall that
we refer to the physics as being effective two-derivative at the horizon if
there is no distinction between the Bekenstein and the Wald entropy densities, see
\eqref{ss2}.} (as in the class of models $\calc_{(2,\Delta)}$), see \eqref{bulks2},
\begin{equation}
\frac{\zeta}{s}\bigg|_{\calb,\calc}=\frac{1}{9\pi} \sum_{i} z_{i,0}^2\,,
\eqlabel{our2}
\end{equation}
where $z_{i,0}$ has to be evaluated at the horizon including $\calo(\beta)$ corrections.

Fig.~\ref{figure6} demonstrates that the naive application of the EO formula \eqref{zeeo} does not work in
higher-derivative models.
By ``naive" application we mean the evaluation of the speed of the sound waves and the background
scalar derivatives at the horizon in \eqref{zeeo} to order $\calo(\beta)$.
Again, the solid curves represent the corrections\footnote{The quasinormal mode analysis \eqref{disprel}
validates these results.} to the
bulk viscosity from \eqref{bulka} (the left panel) and \eqref{bulkbc} (the right panel), and the dashed
green curves indicate corrections from the EO formula \eqref{zeeo}.
Interestingly, there is a disagreement even for models belonging to class $\calb$ (the right panel),
which are effectively two-derivative in the bulk.
One might wonder whether the comparison of the ratio $\frac{\zeta}{s}$,
which is somewhat more universal as it is partly applicable to higher-derivative theories (see \eqref{our2}),
would fare better. From \cite{Eling:2011ms},
\begin{equation}
\frac{\zeta}{s}=c_s^4\cdot \frac{\eta}{s}\cdot \sum_i \left(\frac{\del \phi_{i}}{\del \ln T}\right)^2\bigg|_{\lambda_{\Delta}={\rm const}} \,.
\eqlabel{zetaseo}
\end{equation}
As we show in fig.~\ref{fig6a}, this is not the case for the effective
two-derivative in the bulk model $\calb_{2,3}$: the solid curve represents the
 $\calo(\beta)$ correction $\dd \frac{\zeta}{s}$ extracted from the quasinormal mode \eqref{disprel}
analysis, the red dashed curve is obtained from \eqref{our2}, and the green dashed curve is the
application of the EO formula \eqref{zetaseo}. 
It should probably not come as a surprise that the
EO formula for the bulk viscosity fails in higher derivative as well as in Gauss-Bonnet holographic models:
the naive application of the EO formalism  does not capture the Gauss-Bonnet coupling correction to the shear viscosity either
 \cite{Kats:2007mq}.

In the holographic models $\cala-\calc$ \eqref{dl2} doing a metric field redefinition removes the bulk higher derivative
coupling constants $\alpha_1$ and $\alpha_2$ \cite{Kats:2007mq}. However, in the presence of the scalar sector,
as in \eqref{2der}, such a redefinition generates new higher derivative coupling constants
in the scalar sector of the form $R\cdot (\del\phi)^2$ and $R^{\mu\nu} \del_\mu \phi \del_\nu \phi$.
Because computing the shear viscosity involves only the metric fluctuations, see \eqref{h12},
such a redefinition does not affect it, and the final correction is universal for
all these models, see \eqref{etaac}. On the contrary, computing the bulk viscosity
necessitates turning on the bulk scalar fluctuations, see \eqref{hbulk}. 
As a result, the bulk viscosity is different in models $\cala_{2,\Delta}\,,\, \calb_{2,\Delta}$ and $\calc_{2,\Delta}$
even though all these models have the same value of $\alpha_3=1$.
This is shown explicitly in fig.~\ref{figure7}. Black curves represent $\cala_{2,\Delta}$ models, blue curves
represent $\calb_{2,\Delta}$ models, and magenta curves represent $\calc_{2,\Delta}$ models. 

In fig.~\ref{figure8} we present bulk viscosity corrections in the models $\cald_{4,3}$ (the left panel)
and $\cald_{4,2}$ (the right panel). The solid curves show the bulk viscosity corrections extracted from
the dispersion relation \eqref{disprel}, and the dashed red curves represent \eqref{bulkd}.
This is an excellent validation of our computational framework in holographic models
with Weyl$ ^4$ higher derivative corrections.

We conclude this section  mentioning one of the numerous consistency tests we performed.
The speed of the sound waves $c_s$ can be extracted from the dispersion
relation of the sound channel quasinormal modes of the background black brane
\eqref{disprel}, or from the background black brane equations of state \eqref{cs2}.
We parameterize the speed of the sound waves as
\begin{equation}
c_s=\frac{\beta_{1,0}}{\sqrt{3}}\biggl(1+\beta\cdot \beta_{1,1}\biggr)\,.
\eqlabel{soundd3}
\end{equation}
In fig.~\ref{figure9} we compare results for $(\beta_{1,0})^2$ (the left panel)
and $\beta_{1,1}$ (the right panel). Solid curves indicate data from the dispersion relation
\eqref{disprel}, and the dashed red curves are the corresponding results obtained from the
equation of state \eqref{cs2} in the holographic model $\cald_{4,3}$.

\section{Conclusion}\label{conclude}

In this paper we developed a novel framework for computing transport coefficients in holographic model
with higher derivative corrections. This allowed us to produce compact expressions \eqref{etas2}-\eqref{bulks4}
for the shear and bulk viscosities in large classes of non-conformal holographic models with higher derivative corrections. 
We expect that these formulas would be useful in exploring 
conditions under which  the shear viscosity \cite{Kovtun:2004de} or the bulk viscosity \cite{Buchel:2007mf}  bounds
are violated. The explicit expressions for the Wald entropy density \eqref{ss2} would be useful in searches
of stable holographic conformal order \cite{Chai:2020zgq,Buchel:2020thm,Buchel:2020xdk,Buchel:2020jfs,Buchel:2022zxl}.  
Moreover, since holographic models with scalar fields have been used, depending on the choice of scalar potential, to generate a wide spectrum of temperature dependence for 
$\eta$, in addition to a non-zero $\zeta$, our
analysis is also useful for direct comparison to the physics of the QGP. In particular, 
our simple expressions for the shear and bulk viscosities 
in the presence of arbitrary scalars can facilitate holographic model building and guide the efforts to describe the behavior of the QGP near the deconfinement transition.

We demonstrated that there are particularly simple and universal expressions for the ratio $\frac{\zeta}{s}$,
see \eqref{our2}, valid even in models with higher derivatives in the bulk, but effectively two-derivative physics
at the horizon, specifically when there is no distinction between the Bekenstein and the Wald entropies
of the gauge theory thermal state dual black brane horizon. We also explored the applicability of the Eling-Oz
formula for the bulk viscosity \cite{Eling:2011ms}, and demonstrated that its naive application
fails even in effectively two-derivative holographic Gauss-Bonnet models. 
At this stage it is not clear to us how to extend \cite{Eling:2011ms} to capture theories with higher derivatives
and whether that construction can be generalized in a simple way.

Specific models discussed in section \ref{checks} can be of interest to phenomenological applications
in heavy ion collisions. To facilitate these applications we recall the relations of some of the
parameters used on the gravitational side of the holographic correspondence to the gauge theory
observables:
\begin{itemize}
\item If $c$ and $a$ are the two central charges of a gauge theory UV fixed point,
\begin{equation}
\beta\cdot \alpha_3=\frac{c-a}{8c}\,.
\eqlabel{al3ca}
\end{equation}
The holographic coupling $\alpha_3$ appears in models $\cala-\calc$. We are not
aware of the simple relation for the other two coupling constants, $\alpha_2$ and $\alpha_3$,
in the models $\calb_{2,\Delta}$ and $\calc_{2,\Delta}$.  
\item Assume for simplicity\footnote{See \cite{Buchel:2008ae} for more examples.}
that the UV fixed point is $\caln=4$ $SU(N_c)$
supersymmetric Yang-Mills theory with a gauge coupling $g_{YM}^2$. Then, in models $\cald_{4,\Delta}$,  
\begin{equation}
\beta\equiv \frac 18 \zeta(3)\ \left(g_{YM}^2 N_c\right)^{-3/2}\,.
\eqlabel{dbeta}
\end{equation}
\end{itemize}

In this paper we focused on the first-order transport coefficients, \ie  $\eta$ and $\zeta$, of the
hydrodynamics theory derivative approximation. Stability and causality of the
Landau-frame hydrodynamics can be ensured including higher-order transport coefficients.
Holographic computations of the second-order transport coefficients of conformal gauge theories
were first done in \cite{Baier:2007ix,Bhattacharyya:2007vjd}\footnote{See also \cite{Buchel:2009hv} for some
extensions to nonconformal models.}, and finite 't Hooft coupling corrections
were discussed in \cite{Buchel:2008bz,Buchel:2008kd,Saremi:2011nh,Grozdanov:2016fkt}. 
It would be interesting, albeit challenging, to extend the computational framework proposed
here to the analysis of these coefficients in holographic models with higher derivative corrections.
We expect  that such results will be sensitive to the holographic renormalization of the models,
as well as to the details of the proper formulation of the variational principle,
\ie the precise expressions for the Gibbons-Hawking terms.

In the future, it would also be interesting to extend the results reported here to
holographic models with conserved charges, and to capture the effects of a chemical potential.
It is natural to wonder whether, in the presence of generic higher derivative terms, 
the conductivity can also be extracted from a radially conserved current, and thus entirely from the horizon of the geometry.
Finally, it would be useful to extend our framework to magnetohydrodynamics, again in the
presence of higher derivatives. 
We leave these questions to future work.

\section*{Acknowledgments}
Research at Perimeter
Institute is supported by the Government of Canada through Industry
Canada and by the Province of Ontario through the Ministry of
Research \& Innovation. A.B.'s work was further supported by
NSERC through the Discovery Grants program.
The work of S.C. and L.E. is supported in part by the NSF grant PHY-2210271.
This work was initiated during the
program “The Many Faces of Relativistic Fluid Dynamics”
at the Kavli Institute for Theoretical Physics at UC Santa Barbara.
AB and SC thank KITP for their hospitality. This work was
supported in part by the National Science Foundation under
Grant No. NSF PHY-1748958.

\appendix

\section{Transport coefficients from Kubo formulas}\label{proof}

\subsection{Black brane geometry dual to thermal states of the
boundary theory}\label{background}

The background geometry dual to a thermal equilibrium state
of a boundary gauge theory takes form
\begin{equation}
ds_5^2=-c_1^2\ dt^2+c_2^2\ d\bm{x}^2+ c_3^2\ dr^2\,,
\eqlabel{5metric}
\end{equation}
where $c_i=c_i(r)$, and additionally $\phi_i=\phi_i(r)$.
The radial coordinate is $r\in [0,r_h]$, with $r_h$ being the
location of the regular black brane horizon,
\begin{equation}
\lim_{r\to r_h} c_1=0\,.
\eqlabel{defrh}
\end{equation}
Notice that at this stage we do not fix the residual diffeomorphism
associated with the reparametrization of the radial coordinate. 

One can efficiently compute the background equations of motion
from the effective one dimensional action,
\begin{equation}
S_1=\frac{1}{16\pi G_N}\int_0^{r_h} dr \biggl[\cali+\beta\cdot \dd\cali\biggr]\,,
\eqlabel{s1}
\end{equation}
obtained from the
evaluation of \eqref{2der} on the ansatz \eqref{5metric}.
Here ($'\equiv \frac{d}{dr}$),
\begin{equation}
\cali=c_1 c_2^3 c_3 \biggl(12
-\frac{2 c_1''}{c_1 c_3^2}-\frac{6 c_2''}{c_2 c_3^2}-\frac{6 (c_2')^2}{c_2^2 c_3^2}
+\frac{6 c_3' c_2'}{c_2 c_3^3}-\frac{6 c_1' c_2'}{c_2 c_1 c_3^2}
+\frac{2 c_1' c_3'}{c_1 c_3^3}-\frac{1}{2c_3^2}\sum_i (\phi_i')^2-V
\biggr)\,,
\eqlabel{i2der}
\end{equation}
with the higher derivative contributions in model  \eqref{dl2} given by
\begin{equation}
\begin{split}
&\dd\cali_2=c_1 c_2^3 c_3 \biggl(4
\alpha_1\ \biggl[
\frac{3 c_2''}{c_2 c_3^2}+\frac{c_1''}{c_1 c_3^2}-\frac{c_1' c_3'}{c_1 c_3^3}+
\frac{3c_1' c_2'}{c_2 c_1 c_3^2}-\frac{3 c_3' c_2'}{c_2 c_3^3}
+\frac{3(c_2')^2}{c_2^2 c_3^2}\biggr]^2+\alpha_2\ \biggl[
\frac{6 c_1' c_2'c_2''}{c_2^2 c_1 c_3^4}
\\&-\frac{6 c_1' c_3' c_2''}{c_2 c_1 c_3^5}
-\frac{6 c_2' c_3' c_1''}{c_2 c_1 c_3^5}
-\frac{6 c_1'(c_2,')^2 c_3'}{c_2^2 c_1 c_3^5}
-\frac{24 c_2' c_3' c_2''}{c_2^2 c_3^5}
-\frac{12 (c_2')^3 c_3'}{c_2^3 c_3^5}
+\frac{6 c_1' c_2' (c_3')^2}{c_2 c_1 c_3^6}
+\frac{6 c_1' c_2' c_1''}{c_2 c_1^2 c_3^4}\\
&+\frac{6 c_1''c_2''}{c_2 c_1 c_3^4}
+\frac{12 c_1' (c_2')^3}{c_2^3 c_1 c_3^4}
+\frac{12 (c_2')^2c_2''}{c_2^3 c_3^4}
-\frac{4 c_1' c_3' c_1''}{c_1^2 c_3^5}
-\frac{6 (c_1')^2 c_2' c_3'}{c_2 c_1^2 c_3^5}
+\frac{2 (c_1'')^2}{c_1^2 c_3^4}
+\frac{12 (c_2')^4}{c_2^4 c_3^4}\\&+\frac{2 (c_1')^2 (c_3')^2}{c_1^2 c_3^6}
+\frac{12 (c_1')^2(c_2')^2}{c_2^2 c_1^2 c_3^4}
+\frac{12 (c_2'')^2}{c_2^2 c_3^4}+\frac{12 (c_2')^2 (c_3')^2}{c_2^2 c_3^6}\biggr]
+\alpha_3\ \biggl[\frac{4 (c_1'')^2}{c_1^2 c_3^4}
+\frac{12 (c_1')^2 (c_2')^2}{c_2^2 c_1^2 c_3^4}\\&+\frac{12 (c_2'')^2}{c_2^2 c_3^4}
+\frac{12 (c_2')^4}{c_2^4 c_3^4}-\frac{8 c_1' c_3' c_1''}{c_1^2 c_3^5}
-\frac{24 c_2'c_3' c_2''}{c_2^2 c_3^5}
+\frac{4 (c_1')^2 (c_3')^2}{c_1^2 c_3^6}
+\frac{12 (c_2')^2 (c_3')^2}{c_2^2 c_3^6}\biggr]\,,
\end{split}
\eqlabel{di2}
\end{equation}
and in model \eqref{dl4}  by
\begin{equation}
\begin{split}
&\dd\cali_4=\frac{5}{36} c_1 c_2^3 c_3 \biggl(
\frac{c_1' c_2'}{c_2 c_1 c_3^2}+\frac{c_1' c_3'}{c_1 c_3^3}
-\frac{(c_2')^2}{c_2^2 c_3^2}-\frac{c_3' c_2'}{c_2 c_3^3}
-\frac{c_1''}{c_1 c_3^2}+\frac{c_2''}{c_2 c_3^2}\biggr)^4\,.
\end{split}
\eqlabel{di4}
\end{equation}
From \eqref{s1} we obtain the following equations of
motion\footnote{The $\cdots$ represent the $\calo(\beta)$ terms that
we omit for readability. Of course, these terms must be taken into account
to obtain the correct results.}:
\begin{equation}
\begin{split}
&0=c_1''-\frac{c_1 (c_2')^2}{c_2^2}+\frac{2 c_2' c_1'}{c_2}
-\frac{c_3' c_1'}{c_3}+\frac{1}{12} c_1\ \sum_i(\phi_i')^2+\frac16 c_3^2 c_1 (V-12)+\beta\cdot\biggl[\cdots\biggr]\,,
\end{split}
\eqlabel{eq1}
\end{equation}
\begin{equation}
\begin{split}
&0=c_2''-\frac{c_2' c_3'}{c_3}
+\frac{(c_2')^2}{c_2}+\frac{1}{12} c_2\ \sum_i(\phi_i')^2 +\frac16 c_2 c_3^2 (V-12)+\beta\cdot\biggl[\cdots\biggr]\,,
\end{split}
\eqlabel{eq2}
\end{equation}
\begin{equation}
\begin{split}
&0=\sum_i(\phi_i')^2 -\frac{12 (c_2')^2}{c_2^2}-\frac{12 c_2' c_1'}{c_2 c_1}
-2 c_3^2 (V-12)
+\beta\cdot\biggl[\cdots\biggr]\,,
\end{split}
\eqlabel{eq3}
\end{equation}
\begin{equation}
\begin{split}
&0=\phi_i''-\frac{\phi_i' c_3'}{c_3}+\frac{3 \phi_i' c_2'}{c_2}+\frac{c_1' \phi_i'}{c_1}-c_3^2\ \del_i V\,.
\end{split}
\eqlabel{eq4}
\end{equation}
We verified that the constraint \eqref{eq3} is consistent with the remaining equations to order $\calo(\beta)$
inclusive.

On-shell, \ie evaluated when \eqref{eq1}-\eqref{eq4} hold, the effective action \eqref{s1} is a total derivative. Specifically,
we find 
\begin{equation}
\cali+\beta\cdot\dd\cali=-\frac{6c_2^2c_1}{c_3}\ \cdot\ {\rm eq.}\eqref{eq2}
+\frac{d}{dr}\biggl\{-\frac{2c_2^3 c_1'}{c_3} +\beta\cdot \dd\calb
\biggr\}\,,
\eqlabel{totder}
\end{equation}
with the higher derivative terms $\dd\calb$ given by
\begin{equation}
\begin{split}
&\dd\calb_2=-4 (2 \alpha_1+\alpha_2+2 \alpha_3) \frac{c_2^3 c_1'''}{c_3^3}
-6 (4 \alpha_1+\alpha_2) \frac{c_1 c_2^2 c_2'''}{c_3^3}+12 (4 \alpha_1+\alpha_2) \frac{c_1 (c_2')^3}{c_3^3}
\\&+\frac{2 c_2(c_1')^2}{c_3^5 c_1} \biggl(
24 c_2' c_2 c_3^2 \alpha_1+6 c_2' c_2 c_3^2 \alpha_2
-8 c_3' c_2^2 c_3 \alpha_1-4 c_3' c_2^2 c_3 \alpha_2
-8 c_3' c_2^2 c_3 \alpha_3\biggr) 
+\frac{4 c_2c_1'}{c_3^5} \biggl(
\\&2 c_3'' c_2^2 c_3 \alpha_1+c_3'' c_2^2 c_3 \alpha_2
+2 c_3'' c_2^2 c_3 \alpha_3+12 (c_2')^2 c_3^2 \alpha_1
+3 (c_2')^2 c_3^2 \alpha_2+6 (c_2')^2 c_3^2 \alpha_3
\\&+6 c_2' c_2 c_3 \alpha_1 c_3'+3 c_2' c_2 c_3 \alpha_2 c_3'
+6 c_3' c_2 c_3 \alpha_3 c_2'-6 (c_3')^2 c_2^2 \alpha_1
-3 (c_3')^2 c_2^2 \alpha_2-6 (c_3')^2 c_2^2 \alpha_3\biggr)
\\&-4  (2 \alpha_1+\alpha_2+2 \alpha_3)\frac{c_2^2c_1''}{c_3^4 c_1}
\biggl(
3 c_2' c_3 c_1-3 c_3' c_2 c_1-2 c_2 c_3 c_1' 
\biggr)
+6 (4 \alpha_1+\alpha_2) \frac{c_2 c_1 c_3' (c_2')^2}{c_3^4}
\\&+6 (4 \alpha_1+\alpha_2)  \frac{c_2^2c_1c_2'}{c_3^5}
\biggl(
c_3'' c_3-3 (c_3')^2
\biggr) +6 (4 \alpha_1+\alpha_2)  \frac{c_2''}{c_3^6}\biggl(
-c_2' c_2 c_3^3 c_1+3 c_3' c_2^2 c_3^2 c_1
\biggr)\,,
\end{split}
\eqlabel{defb2}
\end{equation}
and
\begin{equation}
\begin{split}
&\dd\calb_4=\frac59 \biggl(
\frac{(c_2')^2}{c_2^2}+\frac{c_2' c_3'}{c_3 c_2}
-\frac{c_2''}{c_2}-\frac{c_2' c_1'}{c_2 c_1}
-\frac{c_3' c_1'}{c_3 c_1}+\frac{c_1''}{c_1}
\biggr)^2\ \cdot\
\biggl(
\frac{3 c_3'' c_1' c_2^3}{c_3^8}
-\frac{3 c_1 c_3'' c_2' c_2^2}{c_3^8}
-\frac{3 c_1''' c_2^3}{c_3^7}
\\&+\frac{3 c_1 c_2''' c_2^2}{c_3^7}
-\frac{4 (c_1')^2 c_2' c_2^2}{c_3^7 c_1}
-\frac{4 (c_1')^2 c_2^3 c_3'}{c_3^8 c_1}+
\frac{2 c_1' (c_2')^2 c_2}{c_3^7}-\frac{c_1' c_2' c_2^2 c_3'}
{c_3^8}+\frac{4 c_1' c_2^3 c_1''}{c_3^7 c_1}
+\frac{2 c_1' c_2^2 c_2''}{c_3^7}\\
&-\frac{9 c_1' c_2^3 (c_3')^2}{c_3^9}
+\frac{2 c_1 (c_2')^3}{c_3^7}+\frac{5 c_1 (c_2')^2 c_2 c_3'}{c_3^8}
-\frac{c_2' c_2^2 c_1''}{c_3^7}-\frac{5 c_1 c_2' c_2 c_2''}{c_3^7}
+\frac{9 c_1 c_2' c_2^2 (c_3')^2}{c_3^9}
+\frac{9 c_2^3 c_3' c_1''}{c_3^8}\\
&-\frac{9 c_1 c_2^2 c_3'c_2''}{c_3^8}
\biggr)\,.
\end{split}
\eqlabel{defb4}
\end{equation}

In what follows we will need the entropy density $s$ and the temperature $T$ of the boundary thermal state.
The temperature is determined by requiring the vanishing of the conical deficit angle of the 
analytical continuation of the geometry \eqref{5metric},
\begin{equation}
2\pi T = \lim_{r\to r_h}\left[ -\frac{c_2}{c_3}\ \left(\frac{c_1}{c_2}\right)'\right]= \lim_{r\to r_h}
\left[-\frac{c_1'}{c_3}+\frac{c_1c_2'}{c_2 c_3}\right]=\lim_{r\to r_h}
\left[-\frac{c_1'}{c_3}\right]\,,
\eqlabel{deft}
\end{equation}
where to obtain the last equality we used \eqref{defrh}.
The thermal entropy density of the boundary gauge theory is identified with the
entropy density of the dual black brane \cite{Witten:1998zw}. Since our holographic model
contains higher-derivative terms, the Bekenstein entropy $s_B$,
\begin{equation}
s_B=\lim_{r\to r_h} \frac{c_2^3}{4 G_N}\,,
\eqlabel{sb}
\end{equation}
must be replaced with the Wald entropy $s_W$ \cite{Wald:1993nt},
\begin{equation}
s_W=-\frac{1}{8\pi G_N}\lim_{r\to r_h}\biggl[ c_2^3\ \epsilon_{\mu\nu}\epsilon_{\rho\lambda}
\frac{\dd L_5}{\dd R_{\mu\nu\rho\lambda}} \biggr]\,,
\eqlabel{sw}
\end{equation}
\ie $s=s_W$.
The simplest way to compute the Wald entropy density is instead to use the boundary thermodynamics:
\nxt According to the holographic correspondence \cite{Maldacena:1997re,Aharony:1999ti}, the on-shell
gravitational action $S_1$, properly renormalized \cite{Skenderis:2002wp},
has to be identified with the boundary gauge theory free energy density
$\calf$ as follows,
\begin{equation}
-\calf=S_1\bigg|_{\rm on-shell} =\frac{1}{16\pi G_N} \int_0^{r_h}
dr\ \frac{d}{dr}\biggl\{-\frac{2c_2^3 c_1'}{c_3} +\beta\cdot \dd\calb
\biggr\}  +\lim_{r\to 0}\ \biggl[S_{GH}+S_{ct}\biggr]\,,
\eqlabel{deff}
\end{equation}
where we used \eqref{totder}. $S_{GH}$ is a generalized Gibbons-Hawking term \cite{Buchel:2004di},
necessary to have a well-defined variational principle, and $S_{ct}$ is the counter-term
action --- we will not need the explicit form of either of these corrections.  
\nxt Eq.\eqref{deff} can be rearranged to explicitly implement the basic
thermodynamic relation $-\calf=s T-\cale$ between the free energy density $\calf$,
the energy density $\cale$ and the entropy density $s$ \cite{Buchel:2004hw}:
\begin{equation}
-\calf=\frac{1}{16\pi G_N}\lim_{r\to r_h}\biggl[-\frac{2c_2^3 c_1'}{c_3} +\beta\cdot \dd\calb
\biggr] - \lim_{r\to 0}\biggl[\frac{1}{16\pi G_N}\left(
-\frac{2c_2^3 c_1'}{c_3} +\beta\cdot \dd\calb
\right)+S_{GH}+S_{ct}\biggr]\,.               
\eqlabel{btr}
\end{equation}
\nxt Finally, we identify\footnote{Strictly speaking, \eqref{defst} is correct
up to an arbitrary constant. But this constant must be set to zero from the
comparison with thermal AdS, in which case the black brane geometry
is dual to a thermal state of a boundary CFT with vanishing entropy
in the limit $T\to 0$.}
\begin{equation}
s T \equiv  s_W T = \frac{1}{16\pi G_N}\lim_{r\to r_h}\biggl[-\frac{2c_2^3 c_1'}{c_3} +\beta\cdot \dd\calb
\biggr]\,.
\eqlabel{defst}
\end{equation}

Notice that to leading order $\calo(\beta^0)$, using \eqref{deft}, we recover from
\eqref{defst} the Bekenstein entropy
\eqref{sb}; the order $\calo(\beta)$ term implements the correction to get the Wald entropy density.
In what follows we will need the ratio of the Wald and the Bekenstein entropy densities of the
black brane, evaluated at the same temperature:
\begin{equation}
\frac{s_W}{s_B}\equiv 1+\beta\cdot \lim_{r\to r_h} \kappa(r)\,,\qquad \kappa\equiv -\frac{c_3}{2c_2^3c_1'}\cdot
\dd\calb\,.
\eqlabel{sbsw}
\end{equation}
We proceed to evaluate $\kappa$ for our two holographic models \eqref{dl2} and \eqref{dl4}.
Since we work to order $\calo(\beta)$ in \eqref{sbsw}, we can evaluate $\dd \calb$ to leading order
in $\beta$, \ie we can use the leading order equations of motion \eqref{eq1}-\eqref{eq4},
and algebraically eliminate $c_1''', c_2''', c_1'', c_2''$ from $\dd\calb_2$ and $\dd\calb_4$:
\begin{equation}
\begin{split}
&\dd\calb_2 =(16 \alpha_1+5 \alpha_2+4 \alpha_3)\frac{2 c_2^3 c_1} {3c_3} \sum_i \del_iV\cdot  \phi_i'
-c_2^3 \biggl(
(28 \alpha_1+11 \alpha_2+16 \alpha_3) \frac{2c_1'}{3c_3}\\
&+(4 \alpha_1+\alpha_2)\frac{6 c_1 c_2'}{c_3 c_2}
\biggr) V-\frac{36 c_1(c_2')^3}{c_3^3} (4 \alpha_1+\alpha_2)
-\frac{24 c_1' c_2(c_2')^2}{c_3^3} (9 \alpha_1+3 \alpha_2+2 \alpha_3)
\\&-\frac{36 c_2^2 c_2'(c_1')^2}{c_3^3 c_1} (2 \alpha_1+\alpha_2+2 \alpha_3)
+\frac{8 c_2^3c_1'}{c_3} (28 \alpha_1+11 \alpha_2+16 \alpha_3)
+\frac{72 c_2^2 c_2' c_1}{c_3} (4 \alpha_1+\alpha_2)\,,
\end{split}
\eqlabel{b2l}
\end{equation}
\begin{equation}
\begin{split}
&\dd\calb_4 =-\frac{15 (c_2')^2}{c_2^2} \biggl(
\frac{c_1'}{c_1}-\frac{c_2'}{c_2}\biggr)^2 \biggl(
\frac{(c_1' c_2-c_2'c_1)c_2^2}{c_3^5} V
-\frac{12 c_1' c_2^3}{c_3^5}+\frac{12 c_1 c_2' c_2^2}{c_3^5}
+\frac{7 (c_1')^2 c_2' c_2^2}{c_3^7 c_1}\\&+\frac{7 c_1' (c_2')^2 c_2}{c_3^7}
-\frac{14 c_1 (c_2')^3}{c_3^7}
\biggr)\,.
\end{split}
\eqlabel{b4l}
\end{equation}
We need to evaluate \eqref{b2l} and \eqref{b4l} at the horizon, \ie as $r\to r_h$.
It is convenient to fix the residual diffeomorphism in \eqref{5metric}
as\footnote{Of course, final results are independent of this choice.}
\begin{equation}
c_1=\frac{f^{1/2} (g+\calo(\beta))^{1/2}}{r}\,,\qquad c_2=\frac 1r\,,\qquad c_3=\frac{1}{r f^{1/2}(1+\calo(\beta))}\,,
\eqlabel{fixdiff}
\end{equation}
where $\{f,g\}=\{f,g\}(r)$. Given \eqref{eq1}-\eqref{eq3}, to leading order in $\beta$,
\begin{equation}
\begin{split}
&f'=\frac{f}{6r}\ \sum_i (\phi_i')^2+\frac{4f}{r}+\frac{V}{3r}-\frac4r\,,\qquad g'=-\frac{gr}{3}\ \sum_i(\phi_i')^2\,.
\end{split}
\eqlabel{dfdg}
\end{equation}
From \eqref{defrh}, the horizon is located at $r_h$, such that
\begin{equation}
\lim_{r\to r_h} f=0\,.
\eqlabel{rhl}
\end{equation}
Regularity of $\phi_i$ at the horizon then implies from \eqref{eq4} that\footnote{
We use $LHS \sim RHS$ to denote that $\lim_{r\to r_h}\frac{LHS}{RHS}=1$. \label{footnote5}}
\begin{equation}
\phi_i'\ \sim \frac{3\del_iV}{r(V-12)}\,,\qquad {\rm as}\qquad r\to r_h\,.
\eqlabel{dphii}
\end{equation}
Using \eqref{dfdg} and \eqref{dphii} we find from \eqref{b2l}, \eqref{b4l} and \eqref{sbsw}
\begin{equation}
\kappa_2\sim \frac23(V-12)(5\alpha_1+\alpha_2-\alpha_3)\qquad
{\rm and}\qquad \kappa_4\sim -\frac{5}{144}(V-12)^3 \quad {\rm as}\quad r\to r_h\,, 
\eqlabel{defkap}
\end{equation}
\ie
\begin{equation}
\frac{s_W}{s_B}=1+\beta\cdot \begin{cases}
\frac23(V-12)(5\alpha_1+\alpha_2-\alpha_3)\,,\qquad &{\rm when}\ \dd\call=\dd\call_2\,;\\
-\frac{5}{144}(V-12)^3\,,\qquad &{\rm when}\ \dd\call=\dd\call_4\,,
\end{cases}
\eqlabel{ss2}
\end{equation}
where all quantities in \eqref{ss2} are to be evaluated at the black brane horizon.

Finally, note that in the class of models $\dd\call_2$ there are two special cases for which  the
Wald and Bekenstein entropies coincide  to order $\calo(\beta)$:
\begin{itemize}
\item $\alpha_1=1\,,\, \alpha_2=-4\,,\, \alpha_3=1$ --- the combination in \eqref{dl2}
assembles into a Gauss-Bonnet term, which renders the full gravitational action \eqref{2der}
two-derivative;
\item $\alpha_1=0\,,\, \alpha_2=\alpha_3=1$ --- the gravitational action \eqref{2der} is
higher-derivative in the bulk, but is effectively two-derivative in the vicinity of the
black brane horizon.
\end{itemize}
This will be relevant to our discussion later on, when we examine the applicability of the Eling-Oz construction \cite{Eling:2011ms}.

\subsection{$\frac \eta s$ with higher derivative corrections}
\label{etas}

Following \cite{Policastro:2001yc,Buchel:2004di}, we use the Kubo formula to compute the shear
viscosity from the two-point retarded correlation function of the boundary stress-energy tensor
with indices along the spatial directions $ _{12}$ ,
\begin{equation}
\eta=-\lim_{w\to 0}\frac 1w \Im G_R(w)\,,\qquad G_R(w)=-i\int dtd\bm{x} e^{i w t}\theta(t)\langle
[T_{12}(t,{\bm{x}}),T_{12}(0,\bm{0})]\rangle\,.
\eqlabel{kuboeta}
\end{equation}
To compute the retarded thermal two-point function of the components of the stress-energy tensor
entering \eqref{kuboeta} we add the bulk metric perturbation
\begin{equation}
ds_5^2\ \to\ ds_5^2+2 h_{12}(t,r)\ dx_1 dx_2\,.
\eqlabel{h12}
\end{equation}
Simple symmetry arguments \cite{Policastro:2002se} show that all the remaining
metric and bulk scalars fluctuations can be consistently set to zero; additionally,
we can restrict to $SO(3)$ invariant metric perturbations.

It is convenient to
use the idea of the complexified effective action for the fluctuations
introduced in \cite{Gubser:2008sz}. This  complexified action
is a functional of $h_{12,w}(r)$ and $h^*_{12,w}(r)\equiv h_{12,-w}(r)$ 
\begin{equation}
S^{(2)}= \frac{1}{16\pi G_N}\int_0^{r_h} dr\ \call_{\complex}\{h_{12,w},h^*_{12,w}\}\,,
\eqlabel{h12action}
\end{equation}
and is constructed in such a way that the $h_{12,w}$ equation of motion obtained from  $\call_{\complex}$,
\begin{equation}
0=\frac{\dd S^{(2)}}{\dd h^*_{12,w}}\,,
\eqlabel{h12eom}
\end{equation}
is identical to the one obtained from the effective action \eqref{2der}, assuming the
harmonic dependence for the fluctuations $h_{12}(t,r)=e^{- i w t} h_{12,w}(r)$. In practice,
it is straightforward to construct $\call_{\complex}$:
\nxt Evaluate \eqref{2der} to quadratic order in the fluctuations  \eqref{h12}.
\nxt Complexify every term of the resulting quadratic action as follows: \eg replace 
\begin{equation}
\begin{split}
h_{12} \, \del^2_{rr} h_{12}\ &\longrightarrow\ \frac 12 \left(h_{12} \, \del^2_{rr} h_{12}^*+h_{12}^*\, \del^2_{rr} h_{12}\right)\,,
\\ \qquad &{\rm or}\qquad \\
\del_t h_{12} \, \del^2_{tr} h_{12}\ &\longrightarrow\ \frac 12 \left(\del_t h_{12}\, \del^2_{tr} h_{12}^*+\del_t h_{12}^*\, \del^2_{tr}
h_{12}\right)\,.
\end{split}
\eqlabel{complex}
\end{equation}
\nxt Introduce the harmonic dependence as
\begin{equation}
h_{12}=e^{- i w t} h_{12,w}(r)\,,\qquad h_{12}^*=e^{ i w t} h_{12,-w}(r)\,.
\eqlabel{hardep}
\end{equation}
\nxt The resulting action gives $\call_{\complex}$.

For the model \eqref{dl2} we find $\call_{\complex}\equiv \call_{\complex;2}$,
\begin{equation}
\begin{split}
&\call_{\complex;2}=B_1\ h_{12,-w}'' h_{12,w}''
+\frac{B_2}{2}\ \left(h_{12,w}'' h_{12,-w}'
+h_{12,-w}'' h_{12,w}'\right)
\\&-\frac12 (B_3 w^2-A_1-B_4) 
\left( h_{12,w}'' h_{12,-w}
+h_{12,-w}'' h_{12,w}\right)
+(B_{10} w^2+B_{11}+A_4)\ h_{12,w}' h_{12,-w}'
\\&-\frac12 ((B_6-B_9) w^2-A_5-B_{12})\ \left(h_{12,w}' h_{12,-w}+h_{12,-w}' h_{12,w}\right)
\\& +h_{12,w} h_{12,-w}\ \left(B_5 w^4-w^2 (B_7-B_8+A_2-A_3)+B_{13}+A_6\right)\,,
\end{split}
\eqlabel{lc2}
\end{equation}
where the connection coefficients $A_i(r)=\calo(\beta^0)$ and $B_i(r)=\calo(\beta)$ are functionals
of the background geometry \eqref{5metric}. Coefficients $A_i$ are presented in appendix \ref{conshear2};
coefficients $B_i$ are too long to be reported here.
For the model \eqref{dl4} we find $\call_{\complex}\equiv \call_{\complex;4}$ of the form as in
\eqref{lc2}, albeit with a distinct set of the connection coefficients $B_i$.

On-shell, the effective action \eqref{lc2} can be re-expressed as a total derivative,
\begin{equation}
\call_{\complex}=16\pi G_N\cdot h_{12,w}^*\cdot \frac{\dd S^{(2)}}{\dd h^*_{12,w}}+\del_r J_w\,,
\eqlabel{totlc2}
\end{equation}
with a  current 
\begin{equation}
\begin{split}
&J_w=\biggl[B_1 h_{12,w}''+\frac{B_2}{2} h_{12,w}'-\frac12  (B_3 w^2-A_1-B_4)h_{12,w}\biggr] h_{12,-w}'
+\biggl[-B_1 h_{12,w}'''-B_1' h_{12,w}''\\
&+\left(\left(B_{10}+\frac12 B_3\right) w^2-\frac12 B_2'+B_{11}+A_4-\frac12 A_1-\frac12 B_4\right) h_{12,w}'\\
&+\left(\left(-\frac12 B_6+\frac12 B_9+\frac12 B_3'\right) w^2-\frac12 A_1'-\frac12 B_4'+\frac12 A_5+\frac12 B_{12}
\right) h_{12,w}
\biggr] h_{12,-w}\,.
\end{split}
\eqlabel{js}
\end{equation}
A crucial observation  originally made in \cite{Gubser:2008sz}
was that an analog of $J_w$ in two-derivative holographic models
(no $B_i$ coefficients in \eqref{js}) has a radially conserved
imaginary part, on-shell. It is straightforward to verify that this
property holds, even in the presence of higher derivatives of the
effective action:
\begin{equation}
\del_r (J_w-J_{-w})\bigg|_{on-shell}=2\ \del_r \Im J_w\bigg|_{on-shell} =0\,.
\eqlabel{consjw}
\end{equation}
The conservation law \eqref{consjw} is a direct consequence
of the exact $U(1)$ symmetry of $\call_\complex$  \eqref{h12action} that
rotates the phase of  fluctuations, namely $h_{12,w}\to e^{i\theta} h_{12,w}$ and
$h_{12,-w}\to e^{-i\theta} h_{12,-w}$.
The conserved Noether charge associated
with this symmetry is precisely $\Im J_w$.
Indeed, 
\begin{equation}
16\pi G_N\ \frac{\delta S^{(2)}}{\delta \theta} = -i\ \del_r\theta\cdot
(J_w-J_{-w})\,,
\eqlabel{conjw2}
\end{equation}
for infinitesimal $\theta$-rotations. In \cite{Gubser:2008sz}
this conserved charge was interpreted as the radially conserved
number flux of gravitons\footnote{See \cite{Gubser:2008sz}
for further discussion and related earlier work.}$^,$\footnote{It is
an interesting open question as to why the quadratic action
for the fluctuations has this peculiar property, \eqref{totlc2}.}.

Essentially following the discussion of \cite{Son:2002sd}, the retarded two point correlation function of the stress-energy
tensor \eqref{kuboeta} has to be identified with the boundary limit (\ie $r\to 0$) of the current $J_w$ \eqref{js}. 
It is easy to see that $J_w$ diverges as one approaches the AdS boundary --- it must be regularized and renormalized
\cite{Skenderis:2002wp} by adding an appropriate counter-term $J_{ct}$.
Additionally, one must add a generalized Gibbons-Hawking boundary term $J_{GH}$\cite{Buchel:2004di,Cremonini:2009ih} to have a well-defined
variational principle for $S^{(2)}$.
Thus,
\begin{equation}
G_R(w)=\frac{1}{8\pi G_N}\lim_{r\to 0}(J_w+J_{GH}+J_{ct})\,.
\eqlabel{grs}
\end{equation}
Recall from \eqref{kuboeta} that, in order to to extract the shear viscosity, we need only the imaginary part of $G_R(w)$. However, both
$J_{GH}$ and $J_{ct}$ can not contribute to $\Im G_R$. For example, from the representation \eqref{totlc2},
the variation $\frac{\dd S^{(2)}}{\dd h^*_{12,w}}$ would produce a boundary term 
$-A_1 h_{12,w}\cdot \dd h_{12,-w}'$. Such a term must be cancelled with the appropriate term in the variation of $J_{GH}$:
\begin{equation}
\dd\biggl(+A_1\left(h_{12,w}h_{12,-w}'+ h_{12,-w}h_{12,w}'\right) \biggr) = +A_1 h_{12,w}\cdot \dd h_{12,-w}'\,.
\eqlabel{cancel}
\end{equation}
However, $\Im [A_1\left(h_{12,w}h_{12,-w}'+ h_{12,-w}h_{12,w}'\right)]=0$. Clearly, this will be the case for all
terms in $J_{GH}$ and $J_{ct}$: indeed, originally the Gibbons-Hawking term and the counterterms are real, and the complexification
as in \eqref{complex} will not change this fact. As a result,
\begin{equation}
\Im G_R(w) = \frac{1}{8\pi G_N}\lim_{r\to 0} \Im J_w =\frac{1}{8\pi G_N}\lim_{r\to r_h} \Im J_w\,,
\eqlabel{imgrs}
\end{equation}
where in the second equality we used the fact that $\Im J_w$ is conserved along the radial flow, and so \emph{can be evaluated
at the horizon}. Of course, this is the reason underlying the original claims of the universality
of the shear viscosity in two-derivative holographic models \cite{Buchel:2003tz,Buchel:2004qq}. 

Before we can evaluate $J_w$ at the black brane horizon, we need to derive the equation of motion for
$h_{12,w}$. From \eqref{h12eom} we find
\begin{equation}
\begin{split}
&0=(A_1-A_4)\ h_{12,w}''+(A_1'-A_4')\ h_{12,w}'+\left(\frac12 A_1''-\frac12 A_5'+A_6+(A_3-A_2) w^2\right)\ h_{12,w}
\\&+\biggl\{
B_1\ h_{12,w}''''+2 B_1'\ h_{12,w}''+\left(
B_1''+\frac12 B_2'+B_4-B_{11}-w^2 (B_3+B_{10})\right)\ h_{12,w}'\\
&+\biggl(
\frac12 B_2''-B_{11}'+B_4'-(B_3'+B_{10}') w^2\biggr)\ h_{12,w}'+\biggl(
B_5 w^4+\biggl(
B_8-B_7+\frac12 B_6'-\frac12 B_9'\\&-\frac12 B_3''\biggr) w^2+B_{13}+\frac12 B_4''-\frac12 B_{12}'
\biggr)\ h_{12,w}\biggr\}\,.
\end{split}
\eqlabel{eoms}
\end{equation}
All the terms in the bracket $\{\}$ above are $\calo(\beta)$, and the equation \eqref{eoms} can reduced to a second order
equation for $h_{12,w}$ eliminating $h_{12,w}''''\,,\, h_{12,w}'''\,,\, h_{12,w}''$ with $B_i$ connection coefficients
using the $\calo(\beta^0)$ equation of motion.  
We will need to solve the equation \eqref{eoms} in the hydrodynamic approximation, specifically to order $\calo(w)$.
It is convenient to introduce
\begin{equation}
h_{12,w}=c_2^2\ H_{12,w}\,,\qquad H_{12,w}=\left(\frac{c_1}{c_2}\right)^{-i\hw}\left(H_0+i\hw\ H_1\right)\,,
\eqlabel{hydros}
\end{equation}
where we set
\begin{equation}
\hw=\frac{w}{2\pi T}\,.
\eqlabel{defwh}
\end{equation}
With \eqref{hydros}, the incoming wave boundary condition for the fluctuations, and the correct normalization
at the boundary, result in 
\begin{equation}
\begin{split}
&\lim_{r\to 0} H_0(r)=1\,,\qquad \lim_{r\to 0} H_1(r)=0\,, \\
&\lim_{r\to r_h} H_0(r)={\rm finite}\,,\qquad \lim_{r\to r_h} H_1(r)={\rm finite} \,.
\end{split}
\eqlabel{bcshear}
\end{equation}

Using the black brane background equations of motion \eqref{eq1}-\eqref{eq4},  the equation for the fluctuations
\eqref{eoms}, and \eqref{hydros}, we can evaluate the $\calo(\hw)$ part of $J_w$ from \eqref{js}
\begin{equation}
\begin{split}
J_w=J_{0}-i\hw\ J_1\,,\qquad J_1\equiv F+\beta\cdot \dd F\,,
\end{split}
\eqlabel{jwshear}
\end{equation}
\begin{equation}
F=-\frac{c_2^2(c_2 c_1'-c_2' c_1)}{2c_3}\ H_0^2 +\frac{c_2^3 c_1}{2c_3}\ (H_0 H_1'-H_0' H_1)\,,
\eqlabel{fshear}
\end{equation}
while for the model \eqref{dl2}, $\dd F\equiv \dd F_{2}$,
\begin{equation}
\begin{split}
&\dd F_{2}=-\frac23 \biggl(
H_0^2 (c_2 c_1'-c_2' c_1)-c_2 c_1 (H_0 H_1'-H_0' H_1)
\biggr)
\biggl(
\frac{2 c_2^2}{c_3} (2 \alpha_1+\alpha_2+2 \alpha_3) V
\\&+\left(\frac{c_2 c_1' c_2'}{c_1 c_3^3}+\frac{(c_2')^2}{c_3^3}\right) (9 \alpha_1+9 \alpha_2+24 \alpha_3)
+\frac{3 (c_2')^2}{c_3^3} \alpha_3-\frac{24 c_2^2}{c_3} (2 \alpha_1+\alpha_2+2 \alpha_3)
\biggr)\,,
\end{split}
\eqlabel{df2shear}
\end{equation}
and for the model \eqref{dl4}, $\dd F\equiv \dd F_{4}$,
\begin{equation}
\begin{split}
&\dd F_{4}=-\frac{c_2' (c_1 c_2'-c_2 c_1')^2}{4}
\biggl(
H_0^2 (c_2 c_1'-c_2' c_1)-c_2 c_1 (H_0 H_1'-H_0' H_1)
\biggr)
\biggl(
\frac{24 c_2' (c_1')^2}{c_1^4 c_2^2 c_3^7}-\frac{48 c_1'}{c_1^3 c_2 c_3^5}\\&-\frac{96 c_2'}{c_1^2 c_2^2 c_3^5}
+\frac{117 (c_2')^2 c_1'}{c_1^3 c_2^3 c_3^7}
+\frac{111 (c_2')^3}{c_1^2 c_2^4 c_3^7}+\left(\frac{4 c_1'}{c_1^3 c_2 c_3^5}
+\frac{8 c_2'}{c_1^2 c_2^2 c_3^5}\right) V
\biggr)\,.
\end{split}
\eqlabel{df4shear}
\end{equation}
Finally, from \eqref{kuboeta} and \eqref{imgrs} we obtain 
\begin{equation}
\eta=-\frac{1}{8\pi G_N} \lim_{w\to 0} \frac 1w \Im J_w =-\frac{1}{8\pi G_N} \lim_{w\to 0} (-1)\cdot\frac \hw w J_1
=\frac{1}{8\pi G_N}\cdot \frac{1}{2\pi T}\cdot \left(F+\beta\cdot \dd F\right)\,.
\eqlabel{eta} 
\end{equation}

Further simplifications occur when \eqref{fshear}, \eqref{df2shear} and \eqref{df4shear} are evaluated
at the horizon. Using \eqref{fixdiff}-\eqref{dphii}, see also footnote \ref{footnote5}, 
\begin{equation}
\begin{split}
&F\sim (4 G_N s_B)(2\pi T)\cdot \left(\frac 12 H_0^2\right)\,,\\
&\dd F_2\sim (4 G_N s_B)(2\pi T)\cdot \left(\frac {H_0^2}{3}\cdot (V-12)(5\alpha_1+\alpha_2)\right)\,,\\
&\dd F_4\sim (4 G_N s_B)(2\pi T)\cdot \left(-\frac {H_0^2}{96}\cdot (V-12)
\left[5(V-12)^2+2\sum_i\left(\del_iV\right)^2\right]\right)\,,\\
\end{split}
\eqlabel{horequiv}
\end{equation}
where the Bekenstein entropy density is given by \eqref{sb} and the temperature is determined
by \eqref{deft}. 

Notice that in \eqref{horequiv} we only need to know the value of $H_0$ at the black brane
horizon. The equation of motion for $H_0$ is determined from \eqref{eoms} setting $w=0$,
\begin{equation}
0=H_0''+H_0' \left[\ln \frac{c_1c_2^3}{c_3}\right]'+\beta\cdot \dd eq\,,
\eqlabel{eomh0}
\end{equation}
where for the model \eqref{dl2}, $ \dd eq\equiv  \dd eq_2$,
\begin{equation}
\begin{split}
&\dd eq_2=\frac{4H_0'}{3} \biggl[
-\frac{54(c_2')^3}{c_2^3 c_3^2} (\alpha_1+\alpha_2+3 \alpha_3)
-\frac{18 c_1' (c_2')^2}{c_1 c_2^2 c_3^2} (4 \alpha_1+4 \alpha_2+11 \alpha_3)\\
&+\left(\frac{12c_1'}{c_1}-\frac{6 c_2' (c_1')^2}{c_2 c_3^2 c_1^2} \right) (3 \alpha_1+3 \alpha_2+8 \alpha_3)
+\frac{12c_2'}{c_2} (9 \alpha_1+9 \alpha_2+26 \alpha_3)
+2 (2 \alpha_1+\alpha_2\\
&+2 \alpha_3) \sum_i \del_iV\cdot \phi_i'
-\biggl(
(9 \alpha_1+9 \alpha_2+26 \alpha_3) \frac{c_2'}{c_2}+\frac{c_1'}{c_1} (3 \alpha_1+3 \alpha_2+8 \alpha_3)
\biggr) V
\biggr]\,,
\end{split}
\eqlabel{deq2}
\end{equation}
and  for the model \eqref{dl4}, $ \dd eq\equiv  \dd eq_4$,
\begin{equation}
\begin{split}
&\dd eq_4=\frac{H_0'}{6} \biggl(\frac{c_1'}{c_1}-\frac{c_2'}{c_2}\biggr)^2
\biggl[
\frac{12 c_2' (2 c_1 c_2'+c_2 c_1')}{c_1 c_3^4 c_2^2}
\sum_i \del_iV\cdot  \phi_i'-\biggl(
\frac{20 c_2'}{c_3^2 c_2}+\frac{4 c_1'}{c_3^2 c_1}\biggr) V^2
+\biggl(
\frac{480c_2'}{c_3^2 c_2}\\&+\frac{96 c_1'}{c_3^2 c_1}-\frac{849 (c_2')^3}{c_3^4 c_2^3}
-\frac{639 c_1' (c_2')^2}{c_3^4 c_2^2 c_1}
-\frac{96 (c_1')^2 c_2'}{c_3^4 c_2 c_1^2}\biggr) V
-\frac{2880 c_2'}{c_3^2 c_2}-\frac{576 c_1'}{c_3^2 c_1}
+\frac{10188 (c_2')^3}{c_3^4 c_2^3}
\\&+\frac{7668 c_1' (c_2')^2}{c_3^4 c_2^2 c_1}
+\frac{1152 (c_1')^2 c_2'}{c_3^4 c_2 c_1^2}
-\frac{5994 (c_2')^5}{c_2^5 c_3^6}-
\frac{8316 c_1' (c_2')^4}{c_1 c_2^4 c_3^6}
-\frac{3402 (c_1')^2 (c_2')^3}{c_1^2 c_2^3 c_3^6}
\\&-\frac{432 (c_1')^3 (c_2')^2}{c_1^3 c_2^2 c_3^6}
\biggr]\,.
\end{split}
\eqlabel{deq4}
\end{equation}
We seek solution of \eqref{eomh0} recursively in $\beta$,
\begin{equation}
H_0=H_{0,0}+\beta\ H_{0,1} \,.
\eqlabel{h001}
\end{equation}
To leading order,  we have
\begin{equation}
H_{0,0}=\calc_{1,0}+\calc_{2,0} \int dr\ \frac{c_3}{c_2^3 c_1}\,. 
\eqlabel{h00}
\end{equation}
Regularity of $H_{0,0}$ at the horizon fixes $\calc_{2,0}=0$ (see \eqref{fixdiff} and \eqref{rhl}),
while the normalization as $r\to 0$ sets $\calc_{1,0}=1$. With $H_{0,0}\equiv 1$, the general solution
for $H_{0,1}$ once again takes the form \eqref{h00}, albeit now the boundary conditions
\eqref{bcshear} set $\calc_{1,1}=\calc_{2,1}=0$. 
Thus, putting all this together, we have
\begin{equation}
H_0(r)\equiv 1+\calo(\beta^2)\,.
\eqlabel{finh0}
\end{equation}

Finally, collecting \eqref{eta}, \eqref{horequiv} and \eqref{finh0} we find
\begin{equation}
4\pi \frac{\eta}{s_B}=1+\beta\cdot
\begin{cases}
\frac 23 (5\alpha_1+\alpha_2)(V-12),\qquad &{\rm when}\ \dd\call=\dd\call_2\,;\\
-\frac {1}{48}(V-12)
\left[5(V-12)^2+2\sum_i\left(\del_iV\right)^2\right], &{\rm when}\ \dd\call=\dd\call_4\,.
\end{cases}
\eqlabel{etasb}
\end{equation}
Noting that
\begin{equation}
4\pi \frac{\eta}{s}=4\pi \frac{\eta}{s_B}\cdot \frac{s_B}{s_W}\,,
\eqlabel{eatsfin}
\end{equation}
and using \eqref{ss2} we arrive at the results reported in \eqref{etas2} and \eqref{etas4} for the final shear viscosity to entropy ratio in the presence of higher derivative corrections.

\subsection{$\frac \zeta s$ with higher derivative corrections}
\label{zetaeta}

The computation of the bulk viscosity $\zeta$ parallels the discussion of section
\ref{etas}. The starting point is the Kubo formula \cite{Gubser:2008sz}
\begin{equation}
\zeta=-\frac 49\lim_{w\to 0}\frac{1}{w}\Im G_R(w)\,,\qquad G_R(w)=-i\int dtd\bm{x}e^{iw t}
\theta(t)\langle
[\ft 12T_i^{i}(t,{\bm{x}}),\ft 12T_j^j(0,\bm{0})]\rangle\,.
\eqlabel{kubozeta}
\end{equation}
To compute the relevant retarded correlation function we consider the decoupled set
of $SO(3)$ invariant metric fluctuations and the bulk scalars
\begin{equation}
\begin{split}
ds_5^2\ &\to\ ds_5^2+h_{tt}(t,r)\ dt^2+h_{11}(t,r)\ d\bm{x}^2+2 h_{tr}(t,r)\ dtdr+h_{rr}(t,r)\ dr^2\,,\\
\phi_i\ &\to \phi_i+\psi_i(t,r)\,.
\end{split}
\eqlabel{hbulk}
\end{equation}
For convenience, we fix the axial gauge as 
\begin{equation}
h_{tr}=h_{rr}=0\,.
\eqlabel{gaugefix}
\end{equation}
To study the equations of motion for the fluctuations, it is convenient to introduce
new variables $H_{00}\,,\, H_{11}$
\begin{equation}
h_{tt}(t,r)=e^{-i w t}\ c_1^2\ H_{00}(r)\,,\qquad h_{11}(t,r)=e^{-i w t}\ c_2^2\ H_{11}(r)\,,
\eqlabel{redefhb}
\end{equation}
and the gauge invariant scalar fluctuations $Z_i$ as in \cite{Benincasa:2005iv}
\begin{equation}
\psi_i(t,r)=e^{-i wt} \biggl(Z_i(r)+\frac{\phi_i'c_2}{2c_2'}\ H_{11}(r)\biggr) \,.
\eqlabel{defz}
\end{equation}
The equations of motion for $Z_i$ completely decouple from the equations
for the metric fluctuations,
\begin{equation}
\begin{split}
&0=Z_i''+\left(\ln\frac{c_1c_2^3}{c_3}\right)' Z_i' +\frac{c_3^2 w^2}{c_1^2}\ Z_i
-\phi_i'\cdot (V-12)\ \frac{c_2^2 c_3^2}{9(c_2')^2}\ \sum_j Z_j\cdot \phi_j'
\\&-\frac{c_2 c_3^2}{3c_2'} \biggl(
\del_iV\ \sum_j Z_j\cdot \phi_j'+\phi_i'\  \sum_j \del_j V\cdot Z_j
\biggr)
-c_3^2\ \sum_j Z_j\cdot \del^2_{ij}V
+\beta\cdot\biggl[\cdots\biggr]\,.
\end{split}
\eqlabel{zieoms}
\end{equation}
Furthermore, we have
\begin{equation}
\begin{split}
&H_{11}\equiv
\frac{c_2'}{c_2c_3}\ H\,,\qquad  0=H'+\frac{c_3 c_2}{3c_2'}\ \sum_i Z_i\cdot \phi_i'
+\beta\cdot\biggl[\cdots\biggr]\,,
\end{split}
\eqlabel{hbulkv}
\end{equation}
\begin{equation}
\begin{split}
&0=H_{00}'+\frac{c_2}{3c_2'}\ \sum_i\phi_i'\cdot Z_i'
-\frac13 \biggl(
c_3 (V-12)+\frac{3 c_3 w^2}{c_1^2}+\frac{9 c_1' c_2'}{c_3 c_2 c_1}
+\frac{3 (c_1')^2}{c_3 c_1^2}\biggr) H
\\&-\frac{c_2}{9c_1 (c_2')^2}
\biggl(
c_1 c_2 c_3^2 (V-12)+3 c_1' c_2'
\biggr)\ \sum_i Z_i\cdot \phi_i'
-\frac{c_2c_3^2}{3c_2'}\  \sum_i \del_iV\cdot  Z_i
+\beta\cdot\biggl[\cdots\biggr]\,,
\end{split}
\eqlabel{h00bulk}
\end{equation}
where once again we omit the $\calo(\beta)$ terms for readability. 
We will need to solve the equations \eqref{zieoms}-\eqref{h00bulk}
in the hydrodynamic approximation, specifically to order $\calo(w)$
(see \eqref{defwh} for the definition of $\hw$),
\begin{equation}
Z_i=\left(\frac{c_1}{c_2}\right)^{-i\hw}(z_{i,0}+i\hw\ z_{i,1})\,,\qquad H=H_0+i\hw\ H_1\,,\qquad
H_{00}=H_{00,0}+i\hw\ H_{00,1}\,.
\eqlabel{hydroz}
\end{equation}
The set of the gauge invariant scalar equations is solved first. 
The incoming wave boundary condition
implies that $z_{i,0}$ and $z_{i,1}$ must be regular at the black brane horizon. To correctly normalize
the retarded correction function in \eqref{kubozeta} we must set
\begin{equation}
\lim_{r\to 0} H_{11}(r)=1\,,\qquad \lim_{r\to 0} H_{00}(r)=0\,.
\eqlabel{normh11}
\end{equation}
Additionally, the coefficients  of $\psi_i$ that are non-normalizable near the AdS boundary must vanish \cite{Benincasa:2005iv}.
From \eqref{defz} this implies that if the non-normalizable coefficient $\lambda_i$ of the background scalar
$\phi_i$ dual to a gauge theory operator of dimension $\Delta_i$ is nonzero, \ie 
\begin{equation}
\phi_i=\lambda_i\cdot r^{4-\Delta_i}+\cdots\,,\qquad {\rm as}\  r\to 0 \,,
\eqlabel{nonnorm}
\end{equation}
the near-AdS boundary asymptotic of $Z_i$ much be
\begin{equation}
\lim_{r\to 0}\ \frac{Z_i}{r^{4-\Delta_i}}=\frac {4-\Delta_i}{2}\cdot \lambda_i=\lim_{r\to 0}\ \frac{z_{i,0}}{r^{4-\Delta_i}}
\,,\qquad \lim_{r\to 0}\ \frac{z_{i,1}}{r^{4-\Delta_i}}=0\,.
\eqlabel{zibads}
\end{equation}
As we show shortly, we will need only the values of the scalars $z_{i,0}$ near the horizon.
These would have to be determined numerically. 

Parallel to our discussion in section \ref{etas}, we compute the complexified action for the
fluctuations. This action is a functional of $\{h_{00,w}\,,\, h_{11,w}\,,\, p_{i,w}\}$,
\begin{equation}
h_{tt}(t,r)=e^{-i wt}\ h_{00,w}\,,\qquad h_{11}(t,r)=e^{-i wt}\ h_{11,w}\,,\qquad \psi_i(t,r)=e^{-i wt}\ p_{i,w}\,,\qquad 
\eqlabel{stripiw}
\end{equation}
and their complex conjugates:
\begin{equation}
\begin{split}
S^{(2)}=\frac{1}{16\pi G_N}\int_0^{r_h} dr\ \call_\complex\{
h_{00,w}\,,\, h_{11,w}\,,\, p_{i,w},h_{00,w}^*\,,\, h_{11,w}^*\,,\, p_{i,w}^*\}\,,
\end{split}
\eqlabel{s2bulk}
\end{equation}
with 
\begin{equation}
\begin{split}
&\call_\complex=\frac12 A_1 \left(h_{11,w}'' h_{00,-w}+h_{11,-w}'' h_{00,w}\right)
+\frac12 A_2 \left(h_{11,w}'' h_{11,-w}+h_{11,-w}'' h_{11,w}\right)
\\&+\frac12 A_3 \left(h_{00,w}'' h_{11,-w}+h_{00,-w}'' h_{11,w}\right)
+\frac12 A_4 \left(h_{00,w}'' h_{00,-w}+h_{00,-w}'' h_{00,w}\right)
\\&+\frac12 A_5 \left(h_{11,w}' h_{00,-w}'+h_{11,-w}' h_{00,w}'\right)
+\frac12 A_6 \left(h_{11,w}' h_{00,-w}+h_{11,-w}' h_{00,w}\right)\\
&+\frac12 A_7 \left(h_{11,w}' h_{11,-w}+h_{11,-w}' h_{11,w}\right)
+\frac12 A_8 \left(h_{11,-w} h_{00,w}'+h_{11,w} h_{00,-w}'\right)\\
&+\frac12 A_9 \left(h_{00,-w} h_{00,w}'+h_{00,w} h_{00,-w}'\right)
+A_{13} h_{00,-w}' h_{00,w}'
\\&+\frac12 \sum_i A_{17,i}\cdot\left( h_{00,-w} p_{i,w}'+h_{00,w} p_{i,-w}'\right)
+\frac12 \sum_iA_{18,i}\cdot \left( h_{11,-w} p_{i,w}'+h_{11,w} p_{i,-w}'\right)
\\&+A_{19} \sum_i p_{i,w}'p_{i,-w}'+\frac 12 (A_{16}+w^2 (A_{12}-A_{10}))
(h_{00,-w} h_{11,w}+h_{00,w} h_{11,-w})\\&+A_{14} h_{00,w} h_{00,-w}
+(A_{15}-w^2 A_{11}) h_{11,w} h_{11,-w}
+\frac12 \sum_i A_{23,i}\cdot\left (p_{i,w} h_{00,-w}+p_{i,-w} h_{00,w}\right)
\\&+\sum_i (A_{21,i}+w^2 A_{20})\cdot p_{i,w} p_{i,-w}
+\frac12\sum_i A_{22,i}\cdot  (h_{11,-w} p_{i,w}+h_{11,w} p_{i,-w})
\\&+\frac12 \sum_{i\ne j}A_{24,ij} (p_{i,w} p_{j,-w}+p_{i,-w} p_{j,w})+\beta\biggl[\cdots\biggr]\,,
\end{split}
\eqlabel{lcbulk}
\end{equation}
where for readability we suppressed $\calo(\beta)$ terms. The $\calo(\beta^0)$ connection  coefficients
$A_i$ are collected in appendix \ref{conbulk}.

On-shell, the effective action \eqref{lcbulk} can be re-expressed as a
total derivative,
\begin{equation}
\call_\complex=16\pi G_N\cdot \biggl(h_{00,w}^*\cdot \frac{\dd S^{(2)}}{\dd h_{00,w}^*}
+h_{11,w}^*\cdot \frac{\dd S^{(2)}}{\dd h_{11,w}^*}+
p_{i,w}^*\cdot \frac{\dd S^{(2)}}{\dd p_{i,w}^*}\biggr)+\del_r J_w\,,
\eqlabel{totbulk}
\end{equation}
with a  current given by
\begin{equation}
\begin{split}
&J_w=\biggl[\frac{A_4}{2} h_{00,w} +\frac{A_3}{2} h_{11,w}\biggr] h_{00,-w}'
+\biggl[\frac{A_1}{2} h_{00,w}+\frac{A_2}{2} h_{11,w}\biggr] h_{11,-w}'
+\biggl[\frac12 (A_5-A_1) h_{00,w}'
\\&-\frac12 A_2 h_{11,w}'+\frac12 (A_7-A_2') h_{11,w}
+\frac12 (A_6-A_1') h_{00,w}\biggr] h_{11,-w}+\biggl[
\frac12 (A_5-A_3) h_{11,w}'\\&+\left(A_{13}-\frac12 A_4\right) h_{00,w}'
+\frac12 (A_9-A_4') h_{00,w}+\frac12 (A_8-A_3') h_{11,w}\biggr]
h_{00,-w}
+\biggl[
\frac{A_{17,i}}{2} h_{00,w}\\&+\frac{A_{18,i}}{2}  h_{11,w}
+A_{19} p_{i,w}\biggr] p_{i,-w}+\beta\biggl\{\cdots\biggr\}\,.
\end{split}
\eqlabel{jbulk}
\end{equation}
Parallel to the discussion of the shear channel fluctuations
in section \ref{etas}, the imaginary part of $J_w$ in \eqref{jbulk}
is radially conserved. 
The same arguments as in section \ref{etas} lead to 
\begin{equation}
\Im G_R(w) =\frac{1}{8\pi G_N}\lim_{r\to r_h} \Im J_w\,,
\eqlabel{imgrb}
\end{equation}
with (see \eqref{hydroz})
\begin{equation}
\begin{split}
J_w=J_{0}-i\hw\ J_1\,,\qquad J_1\equiv F+\beta\cdot \dd F\,,
\end{split}
\eqlabel{jwbulk}
\end{equation}
\begin{equation}
F=-\frac{c_2^2(c_2 c_1'-c_2' c_1)}{2c_3}\ \sum_i (z_{i,0})^2+\frac{c_2^3 c_1}{2c_3}\
\sum_{i} \left(z_{i,1}'z_{i,0}-z_{i,0}'z_{i,1}\right)\,,
\eqlabel{fbulk}
\end{equation}
where we used the equations of motion \eqref{hbulkv} and \eqref{h00bulk}
to eliminate all derivatives of $H_{11}$ and $H_{00}$. What is remarkable is that
the final result \eqref{fbulk} depends only on $z_{i,0}$ and  $z_{i,1}$. 
While we will not present here the results for $\dd F$'s, we note that they 
are  functionals of $z_{i,0}$ and  $z_{i,1}$ only as well. 

As in section \ref{etas}, further simplification occurs when $J_1$ is evaluated at the horizon.
We find:
\begin{equation}
\begin{split}
&F\sim (4 G_N s_B)(2\pi T)\cdot \left(\frac 12 \sum_i (z_{i,0})^2\right)\,,\\
&\dd F_2\sim (4 G_N s_B)(2\pi T)\cdot
\left(\frac{2(5 \alpha_1+\alpha_2-\alpha_3)}{3(V-12)}\ \sum_i (z_{i,0}\cdot\del_i V)^2\right)\,,\\
&\dd F_4\sim (4 G_N s_B)(2\pi T)\cdot \left(-\frac{5}{48}(V-12)\ \sum_i (z_{i,0}\cdot\del_i V)^2
\right)\,,\\
\end{split}
\eqlabel{horequivbulk}
\end{equation}
where the Bekenstein entropy density is given by \eqref{sb} and the temperature is determined
by \eqref{deft}. As before, we used $\dd F \equiv \dd F_2$ to refer to the model \eqref{dl2},
and  $\dd F \equiv \dd F_4$ to refer to the model \eqref{dl4}.

Finally, using \eqref{imgrb} and \eqref{horequivbulk}, we obtain  from \eqref{kubozeta}
the following expressions for the bulk viscosity,
\begin{equation}
\zeta=\frac{s_B}{4\pi}\cdot \frac 49
\biggl[\qquad \sum_i z_{i,0}^2+\beta\cdot\begin{cases}
\frac{4(5 \alpha_1+\alpha_2-\alpha_3)} {3(V-12)}\ \sum_i (z_{i,0}\cdot\del_i V)^2\,,\qquad &{\rm when}\ \dd\call=\dd\call_2\,;\\
-\frac{5}{24}(V-12)\ \sum_i (z_{i,0}\cdot\del_i V)^2\,,\qquad &{\rm when}\ \dd\call=\dd\call_4
\end{cases}
\qquad \biggr] \,.
\eqlabel{zetasb}
\end{equation}
Noting that
\begin{equation}
9\pi \frac{\zeta}{s}=9\pi \frac{\zeta}{s_B}\cdot \frac{s_B}{s_W}\,,
\eqlabel{zetasfin}
\end{equation}
and using \eqref{ss2}, we arrive at the results reported in \eqref{bulks2} and \eqref{bulks4}.

\section{Connection coefficients of \eqref{lc2}}\label{conshear2}

\begin{equation}
A_1=\frac{2 c_1}{c_2 c_3}\,,
\eqlabel{a1s}
\end{equation}
\begin{equation}
A_2=-\frac{2 c_3}{c_1 c_2}\,,
\eqlabel{a2s}
\end{equation}
\begin{equation}
A_3=-\frac{3c_3}{2c_1 c_2}\,,
\eqlabel{a3s}
\end{equation}
\begin{equation}
A_4= \frac{3c_1}{2c_2 c_3}\,,
\eqlabel{a4s}
\end{equation}
\begin{equation}
A_5=-\frac{2 c_1 c_3'}{c_2 c_3^2}-\frac{6 c_1 c_2'}{c_2^2 c_3}
+\frac{2 c_1'}{c_2 c_3}\,,
\eqlabel{a5s}
\end{equation}
\begin{equation}
\begin{split}
&A_6=\frac{c_3c_1}{2c_2}\ V+\frac{c_1}{4c_2c_3}\ \sum_i(\phi_i')^2
-\frac{6 c_3 c_1}{c_2}-\frac{c_2'' c_1}{c_2^2 c_3}
+\frac{c_3' c_2' c_1}{c_2^2 c_3^2}
+\frac{5 (c_2')^2 c_1}{c_2^3 c_3}+\frac{c_1''}{c_2 c_3}
\\&-\frac{c_3' c_1'}{c_2 c_3^2}-\frac{c_1' c_2'}{c_2^2 c_3}\,.
\end{split}
\eqlabel{a6s}
\end{equation}

\section{Connection coefficients of \eqref{lcbulk}}\label{conbulk}

\begin{equation}
A_1=\frac{3c_2}{2c_3 c_1}\,,
\eqlabel{a1b}
\end{equation}
\begin{equation}
A_2=-\frac{3c_1}{2c_2 c_3}\,,
\eqlabel{a2b}
\end{equation}
\begin{equation}
A_3=\frac{3c_2}{2c_3 c_1}\,,
\eqlabel{a3b}
\end{equation}
\begin{equation}
A_4=\frac{c_2^3}{2c_3 c_1^3}\,,
\eqlabel{a4b}
\end{equation}
\begin{equation}
A_5=\frac{3c_2}{2c_3 c_1}\,,
\eqlabel{a5b}
\end{equation}
\begin{equation}
A_6=-\frac{3c_2 c_1'}{2c_1^2 c_3}-\frac{3c_2 c_3'}{2c_1 c_3^2}\,,
\eqlabel{a6b}
\end{equation}
\begin{equation}
A_7=-\frac{3c_1'}{2c_2 c_3}+\frac{3c_1 c_3'}{2c_2 c_3^2}\,,
\eqlabel{a7b}
\end{equation}
\begin{equation}
A_8=-\frac{3 c_2 c_1'}{c_1^2 c_3}-\frac{3c_2 c_3'}{2c_1 c_3^2}+\frac{3c_2'}{2c_1 c_3}\,,
\eqlabel{a8b}
\end{equation}
\begin{equation}
A_9=-\frac{3 c_2^3 c_1'}{c_1^4 c_3}-\frac{c_2^3 c_3'}{2c_1^3 c_3^2}
+\frac{3c_2^2 c_2'}{2c_1^3 c_3}\,,
\eqlabel{a9b}
\end{equation}
\begin{equation}
A_{10}= \frac{3c_2 c_3}{2c_1^3}\,,
\eqlabel{a10b}
\end{equation}
\begin{equation}
A_{11}= \frac{3c_3}{2c_2 c_1}\,,
\eqlabel{a11b}
\end{equation}
\begin{equation}
A_{12}= \frac{3c_2 c_3}{2c_1^3}\,,
\eqlabel{a12b}
\end{equation}
\begin{equation}
A_{13}= \frac{c_2^3}{2c_3 c_1^3}\,,
\eqlabel{a13b}
\end{equation}
\begin{equation}
\begin{split}
&A_{14}=\frac{c_2^3 c_3}{8c_1^3}\ V
+\frac{c_2^3}{16c_3 c_1^3} \sum_i(\phi_i')^2-\frac{3c_2^3 c_3}{2c_1^3}
+\frac{3c_2^2 c_2''}{4c_3 c_1^3}
-\frac{3c_2^3 c_1''}{4c_3 c_1^4}
+\frac{3 c_2^3 (c_1')^2}{c_3 c_1^5}
+\frac{3c_2^3 c_1' c_3'}{4c_3^2 c_1^4}
\\&-\frac{9c_2^2c_1'c_2'}{4c_3 c_1^4}
- \frac{3c_2^2 c_2' c_3'}{4c_3^2 c_1^3}+\frac{3c_2 (c_2')^2}{4c_3 c_1^3}\,,
\end{split}
\eqlabel{a14b}
\end{equation}
\begin{equation}
\begin{split}
&A_{15}=-\frac{3c_3c_1}{8c_2}\ V-\frac{3c_1}{16c_3c_2}\sum_i (\phi_i')^2
+\frac{9c_3 c_1}{2c_2}+\frac{3c_2'' c_1}{4c_3 c_2^2}
-\frac{3c_1''}{4c_3 c_2}
+\frac{3c_1'c_3'}{4c_3^2 c_2}
+\frac{3c_1'c_2'}{4c_3 c_2^2}
\\&- \frac{3c_1 c_2'c_3'}{4c_3^2 c_2^2}
+\frac{3c_1 (c_2')^2}{4c_3 c_2^3}\,,
\end{split}
\eqlabel{a15b}
\end{equation}
\begin{equation}
\begin{split}
&A_{16}=\frac{3c_2 c_3}{4c_1}\ V+\frac{3c_2}{8c_1c_3}\sum_i (\phi_i')^2
-\frac{9 c_2 c_3}{c_1}+\frac{3 c_2''}{2c_1 c_3}
-\frac{3c_2 c_1''}{2c_1^2 c_3}
+\frac{3 c_2 (c_1')^2}{c_1^3 c_3}
+\frac{3c_2 c_1'c_3'}{2c_1^2 c_3^2}
\\&- \frac{3c_1'c_2'}{2c_1^2 c_3}
-\frac{3c_2'c_3'}{2c_1 c_3^2}
+\frac{3(c_2')^2}{2c_2 c_1 c_3}\,,
\end{split}
\eqlabel{a16b}
\end{equation}
\begin{equation}
A_{17,i}=\frac{c_2^3}{2c_3 c_1}\ \phi_i'\,,
\eqlabel{a17b}
\end{equation}
\begin{equation}
A_{18,i}=-\frac{3c_2 c_1}{2c_3}\ \phi_i'\,,
\eqlabel{a18b}
\end{equation}
\begin{equation}
A_{19}=- \frac{c_1 c_2^3}{2c_3}\,,
\eqlabel{a19b}
\end{equation}
\begin{equation}
A_{20}=\frac{c_2^3 c_3}{2c_1}\,,
\eqlabel{a20b}
\end{equation}
\begin{equation}
A_{21,i}=-\frac{c_2^3 c_3 c_1}{2}\ \del^2_{ii}V\,,
\eqlabel{a21b}
\end{equation}
\begin{equation}
A_{22,i}=-\frac{3c_2 c_3 c_1}{2}\ \del_iV\,,
\eqlabel{a22b}
\end{equation}
\begin{equation}
A_{23,i}=\frac{c_3c_2^3}{2c_1}\ \del_iV\,,
\eqlabel{a23b}
\end{equation}
\begin{equation}
A_{24,ij}=-c_2^3 c_3 c_1\ \del^2_{ij}V \,.
\eqlabel{a24b}
\end{equation}

\bibliographystyle{JHEP}
\bibliography{hdt}

\providecommand{\href}[2]{#2}\begingroup\raggedright\begin{thebibliography}{10}

\bibitem{Casalderrey-Solana:2011dxg}
J.~Casalderrey-Solana, H.~Liu, D.~Mateos, K.~Rajagopal and U.~A. Wiedemann,
  \emph{{Gauge/String Duality, Hot QCD and Heavy Ion Collisions}}.
\newblock Cambridge University Press, 2014,
  \href{http://dx.doi.org/10.1017/CBO9781139136747}{10.1017/CBO9781139136747}.

\bibitem{Policastro:2001yc}
G.~Policastro, D.~T. Son and A.~O. Starinets, \emph{{The Shear viscosity of
  strongly coupled N=4 supersymmetric Yang-Mills plasma}},
  \href{http://dx.doi.org/10.1103/PhysRevLett.87.081601}{\emph{Phys. Rev.
  Lett.} {\bf 87} (2001) 081601},
  [\href{https://arxiv.org/abs/hep-th/0104066}{{\tt hep-th/0104066}}].

\bibitem{Buchel:2003tz}
A.~Buchel and J.~T. Liu, \emph{{Universality of the shear viscosity in
  supergravity}},
  \href{http://dx.doi.org/10.1103/PhysRevLett.93.090602}{\emph{Phys. Rev.
  Lett.} {\bf 93} (2004) 090602},
  [\href{https://arxiv.org/abs/hep-th/0311175}{{\tt hep-th/0311175}}].

\bibitem{Baggioli:2023yvc}
M.~Baggioli, S.~Cremonini, L.~Early, L.~Li and H.-T. Sun, \emph{{Breaking
  rotations without violating the KSS viscosity bound}},
  \href{http://dx.doi.org/10.1007/JHEP07(2023)016}{\emph{JHEP} {\bf 07} (2023)
  016}, [\href{https://arxiv.org/abs/2304.01807}{{\tt 2304.01807}}].

\bibitem{Kovtun:2003wp}
P.~Kovtun, D.~T. Son and A.~O. Starinets, \emph{{Holography and hydrodynamics:
  Diffusion on stretched horizons}},
  \href{http://dx.doi.org/10.1088/1126-6708/2003/10/064}{\emph{JHEP} {\bf 10}
  (2003) 064}, [\href{https://arxiv.org/abs/hep-th/0309213}{{\tt
  hep-th/0309213}}].

\bibitem{Kovtun:2004de}
P.~Kovtun, D.~T. Son and A.~O. Starinets, \emph{{Viscosity in strongly
  interacting quantum field theories from black hole physics}},
  \href{http://dx.doi.org/10.1103/PhysRevLett.94.111601}{\emph{Phys. Rev.
  Lett.} {\bf 94} (2005) 111601},
  [\href{https://arxiv.org/abs/hep-th/0405231}{{\tt hep-th/0405231}}].

\bibitem{Cremonini:2011iq}
S.~Cremonini, \emph{{The Shear Viscosity to Entropy Ratio: A Status Report}},
  \href{http://dx.doi.org/10.1142/S0217984911027315}{\emph{Mod. Phys. Lett. B}
  {\bf 25} (2011) 1867--1888}, [\href{https://arxiv.org/abs/1108.0677}{{\tt
  1108.0677}}].

\bibitem{Buchel:2004di}
A.~Buchel, J.~T. Liu and A.~O. Starinets, \emph{{Coupling constant dependence
  of the shear viscosity in N=4 supersymmetric Yang-Mills theory}},
  \href{http://dx.doi.org/10.1016/j.nuclphysb.2004.11.055}{\emph{Nucl. Phys. B}
  {\bf 707} (2005) 56--68}, [\href{https://arxiv.org/abs/hep-th/0406264}{{\tt
  hep-th/0406264}}].

\bibitem{Kats:2007mq}
Y.~Kats and P.~Petrov, \emph{{Effect of curvature squared corrections in AdS on
  the viscosity of the dual gauge theory}},
  \href{http://dx.doi.org/10.1088/1126-6708/2009/01/044}{\emph{JHEP} {\bf 01}
  (2009) 044}, [\href{https://arxiv.org/abs/0712.0743}{{\tt 0712.0743}}].

\bibitem{Brigante:2008gz}
M.~Brigante, H.~Liu, R.~C. Myers, S.~Shenker and S.~Yaida, \emph{{The Viscosity
  Bound and Causality Violation}},
  \href{http://dx.doi.org/10.1103/PhysRevLett.100.191601}{\emph{Phys. Rev.
  Lett.} {\bf 100} (2008) 191601}, [\href{https://arxiv.org/abs/0802.3318}{{\tt
  0802.3318}}].

\bibitem{Buchel:2010wf}
A.~Buchel and S.~Cremonini, \emph{{Viscosity Bound and Causality in Superfluid
  Plasma}}, \href{http://dx.doi.org/10.1007/JHEP10(2010)026}{\emph{JHEP} {\bf
  10} (2010) 026}, [\href{https://arxiv.org/abs/1007.2963}{{\tt 1007.2963}}].

\bibitem{Cremonini:2012ny}
S.~Cremonini, U.~Gursoy and P.~Szepietowski, \emph{{On the Temperature
  Dependence of the Shear Viscosity and Holography}},
  \href{http://dx.doi.org/10.1007/JHEP08(2012)167}{\emph{JHEP} {\bf 08} (2012)
  167}, [\href{https://arxiv.org/abs/1206.3581}{{\tt 1206.3581}}].

\bibitem{Rougemont:2023gfz}
R.~Rougemont, J.~Grefa, M.~Hippert, J.~Noronha, J.~Noronha-Hostler, I.~Portillo
  et~al., \emph{{Hot QCD Phase Diagram From Holographic
  Einstein-Maxwell-Dilaton Models}},
  \href{https://arxiv.org/abs/2307.03885}{{\tt 2307.03885}}.

\bibitem{Parnachev:2005hh}
A.~Parnachev and A.~Starinets, \emph{{The Silence of the little strings}},
  \href{http://dx.doi.org/10.1088/1126-6708/2005/10/027}{\emph{JHEP} {\bf 10}
  (2005) 027}, [\href{https://arxiv.org/abs/hep-th/0506144}{{\tt
  hep-th/0506144}}].

\bibitem{Benincasa:2005iv}
P.~Benincasa, A.~Buchel and A.~O. Starinets, \emph{{Sound waves in strongly
  coupled non-conformal gauge theory plasma}},
  \href{http://dx.doi.org/10.1016/j.nuclphysb.2005.11.005}{\emph{Nucl. Phys. B}
  {\bf 733} (2006) 160--187}, [\href{https://arxiv.org/abs/hep-th/0507026}{{\tt
  hep-th/0507026}}].

\bibitem{Buchel:2005cv}
A.~Buchel, \emph{{Transport properties of cascading gauge theories}},
  \href{http://dx.doi.org/10.1103/PhysRevD.72.106002}{\emph{Phys. Rev. D} {\bf
  72} (2005) 106002}, [\href{https://arxiv.org/abs/hep-th/0509083}{{\tt
  hep-th/0509083}}].

\bibitem{Buchel:2007mf}
A.~Buchel, \emph{{Bulk viscosity of gauge theory plasma at strong coupling}},
  \href{http://dx.doi.org/10.1016/j.physletb.2008.03.069}{\emph{Phys. Lett.}
  {\bf B663} (2008) 286--289}, [\href{https://arxiv.org/abs/0708.3459}{{\tt
  0708.3459}}].

\bibitem{Gubser:2008sz}
S.~S. Gubser, S.~S. Pufu and F.~D. Rocha, \emph{{Bulk viscosity of strongly
  coupled plasmas with holographic duals}},
  \href{http://dx.doi.org/10.1088/1126-6708/2008/08/085}{\emph{JHEP} {\bf 08}
  (2008) 085}, [\href{https://arxiv.org/abs/0806.0407}{{\tt 0806.0407}}].

\bibitem{Buchel:2008uu}
A.~Buchel and C.~Pagnutti, \emph{{Bulk viscosity of N=2* plasma}},
  \href{http://dx.doi.org/10.1016/j.nuclphysb.2009.02.022}{\emph{Nucl. Phys. B}
  {\bf 816} (2009) 62--72}, [\href{https://arxiv.org/abs/0812.3623}{{\tt
  0812.3623}}].

\bibitem{Gursoy:2009kk}
U.~Gursoy, E.~Kiritsis, G.~Michalogiorgakis and F.~Nitti, \emph{{Thermal
  Transport and Drag Force in Improved Holographic QCD}},
  \href{http://dx.doi.org/10.1088/1126-6708/2009/12/056}{\emph{JHEP} {\bf 12}
  (2009) 056}, [\href{https://arxiv.org/abs/0906.1890}{{\tt 0906.1890}}].

\bibitem{Buchel:2011wx}
A.~Buchel, U.~Gursoy and E.~Kiritsis, \emph{{Holographic bulk viscosity: GPR
  versus EO}}, \href{http://dx.doi.org/10.1007/JHEP09(2011)095}{\emph{JHEP}
  {\bf 09} (2011) 095}, [\href{https://arxiv.org/abs/1104.2058}{{\tt
  1104.2058}}].

\bibitem{Buchel:2011uj}
A.~Buchel, \emph{{Violation of the holographic bulk viscosity bound}},
  \href{http://dx.doi.org/10.1103/PhysRevD.85.066004}{\emph{Phys. Rev. D} {\bf
  85} (2012) 066004}, [\href{https://arxiv.org/abs/1110.0063}{{\tt
  1110.0063}}].

\bibitem{Eling:2011ms}
C.~Eling and Y.~Oz, \emph{{A Novel Formula for Bulk Viscosity from the Null
  Horizon Focusing Equation}},
  \href{http://dx.doi.org/10.1007/JHEP06(2011)007}{\emph{JHEP} {\bf 06} (2011)
  007}, [\href{https://arxiv.org/abs/1103.1657}{{\tt 1103.1657}}].

\bibitem{Buchel:2011yv}
A.~Buchel, \emph{{On Eling-Oz formula for the holographic bulk viscosity}},
  \href{http://dx.doi.org/10.1007/JHEP05(2011)065}{\emph{JHEP} {\bf 05} (2011)
  065}, [\href{https://arxiv.org/abs/1103.3733}{{\tt 1103.3733}}].

\bibitem{Gursoy:2007cb}
U.~Gursoy and E.~Kiritsis, \emph{{Exploring improved holographic theories for
  QCD: Part I}},
  \href{http://dx.doi.org/10.1088/1126-6708/2008/02/032}{\emph{JHEP} {\bf 02}
  (2008) 032}, [\href{https://arxiv.org/abs/0707.1324}{{\tt 0707.1324}}].

\bibitem{Gursoy:2007er}
U.~Gursoy, E.~Kiritsis and F.~Nitti, \emph{{Exploring improved holographic
  theories for QCD: Part II}},
  \href{http://dx.doi.org/10.1088/1126-6708/2008/02/019}{\emph{JHEP} {\bf 02}
  (2008) 019}, [\href{https://arxiv.org/abs/0707.1349}{{\tt 0707.1349}}].

\bibitem{Starinets:2008fb}
A.~O. Starinets, \emph{{Quasinormal spectrum and the black hole membrane
  paradigm}},
  \href{http://dx.doi.org/10.1016/j.physletb.2008.11.028}{\emph{Phys. Lett. B}
  {\bf 670} (2009) 442--445}, [\href{https://arxiv.org/abs/0806.3797}{{\tt
  0806.3797}}].

\bibitem{Damour:1979wya}
T.~Damour, \emph{{Quelques proprietes mecaniques, electromagnet iques,
  thermodynamiques et quantiques des trous noir}}.
\newblock PhD thesis, Paris U., VI-VII, 1979.

\bibitem{Thorne:1986iy}
K.~S. Thorne, R.~H. Price and D.~A. Macdonald, eds., \emph{{BLACK HOLES: THE
  MEMBRANE PARADIGM}}.
\newblock 1986.

\bibitem{Iqbal:2008by}
N.~Iqbal and H.~Liu, \emph{{Universality of the hydrodynamic limit in AdS/CFT
  and the membrane paradigm}},
  \href{http://dx.doi.org/10.1103/PhysRevD.79.025023}{\emph{Phys. Rev. D} {\bf
  79} (2009) 025023}, [\href{https://arxiv.org/abs/0809.3808}{{\tt
  0809.3808}}].

\bibitem{Demircik:2023lsn}
T.~Demircik, D.~Gallegos, U.~G\"ursoy, M.~J\"arvinen and R.~Lier, \emph{{A
  Novel Method for Holographic Transport}},
  \href{https://arxiv.org/abs/2311.00042}{{\tt 2311.00042}}.

\bibitem{Donos:2022uea}
A.~Donos, P.~Kailidis and C.~Pantelidou, \emph{{Holographic dissipation from
  the symplectic current}},
  \href{http://dx.doi.org/10.1007/JHEP10(2022)058}{\emph{JHEP} {\bf 10} (2022)
  058}, [\href{https://arxiv.org/abs/2208.05911}{{\tt 2208.05911}}].

\bibitem{Davison:2022vqh}
R.~A. Davison, B.~Gout\'eraux and E.~Mefford, \emph{{Zero sound and higher-form
  symmetries in compressible holographic phases}},
  \href{http://dx.doi.org/10.1007/JHEP12(2023)040}{\emph{JHEP} {\bf 12} (2023)
  040}, [\href{https://arxiv.org/abs/2210.14802}{{\tt 2210.14802}}].

\bibitem{Donos:2022www}
A.~Donos and P.~Kailidis, \emph{{Dissipative effects in finite density
  holographic superfluids}},
  \href{http://dx.doi.org/10.1007/JHEP11(2022)053}{\emph{JHEP} {\bf 11} (2022)
  053}, [\href{https://arxiv.org/abs/2209.06893}{{\tt 2209.06893}}].

\bibitem{Benincasa:2005qc}
P.~Benincasa and A.~Buchel, \emph{{Transport properties of N=4 supersymmetric
  Yang-Mills theory at finite coupling}},
  \href{http://dx.doi.org/10.1088/1126-6708/2006/01/103}{\emph{JHEP} {\bf 01}
  (2006) 103}, [\href{https://arxiv.org/abs/hep-th/0510041}{{\tt
  hep-th/0510041}}].

\bibitem{Buchel:2008ac}
A.~Buchel, \emph{{Shear viscosity of boost invariant plasma at finite
  coupling}},
  \href{http://dx.doi.org/10.1016/j.nuclphysb.2008.03.009}{\emph{Nucl. Phys. B}
  {\bf 802} (2008) 281--306}, [\href{https://arxiv.org/abs/0801.4421}{{\tt
  0801.4421}}].

\bibitem{Buchel:2008sh}
A.~Buchel, \emph{{Resolving disagreement for eta/s in a CFT plasma at finite
  coupling}},
  \href{http://dx.doi.org/10.1016/j.nuclphysb.2008.05.024}{\emph{Nucl. Phys. B}
  {\bf 803} (2008) 166--170}, [\href{https://arxiv.org/abs/0805.2683}{{\tt
  0805.2683}}].

\bibitem{Buchel:2018ttd}
A.~Buchel, \emph{{Non-conformal holographic Gauss-Bonnet hydrodynamics}},
  \href{http://dx.doi.org/10.1007/JHEP03(2018)037}{\emph{JHEP} {\bf 03} (2018)
  037}, [\href{https://arxiv.org/abs/1801.06165}{{\tt 1801.06165}}].

\bibitem{Kovtun:2005ev}
P.~K. Kovtun and A.~O. Starinets, \emph{{Quasinormal modes and holography}},
  \href{http://dx.doi.org/10.1103/PhysRevD.72.086009}{\emph{Phys. Rev. D} {\bf
  72} (2005) 086009}, [\href{https://arxiv.org/abs/hep-th/0506184}{{\tt
  hep-th/0506184}}].

\bibitem{Buchel:2007vy}
A.~Buchel, S.~Deakin, P.~Kerner and J.~T. Liu, \emph{{Thermodynamics of the
  N=2* strongly coupled plasma}},
  \href{http://dx.doi.org/10.1016/j.nuclphysb.2007.06.019}{\emph{Nucl. Phys.}
  {\bf B784} (2007) 72--102}, [\href{https://arxiv.org/abs/hep-th/0701142}{{\tt
  hep-th/0701142}}].

\bibitem{Chai:2020zgq}
N.~Chai, S.~Chaudhuri, C.~Choi, Z.~Komargodski, E.~Rabinovici and M.~Smolkin,
  \emph{{Thermal Order in Conformal Theories}},
  \href{http://dx.doi.org/10.1103/PhysRevD.102.065014}{\emph{Phys. Rev. D} {\bf
  102} (2020) 065014}, [\href{https://arxiv.org/abs/2005.03676}{{\tt
  2005.03676}}].

\bibitem{Buchel:2020thm}
A.~Buchel, \emph{{Thermal order in holographic CFTs and no-hair theorem
  violation in black branes}},
  \href{http://dx.doi.org/10.1016/j.nuclphysb.2021.115425}{\emph{Nucl. Phys. B}
  {\bf 967} (2021) 115425}, [\href{https://arxiv.org/abs/2005.07833}{{\tt
  2005.07833}}].

\bibitem{Buchel:2020xdk}
A.~Buchel, \emph{{SUGRA/Strings like to be bald}},
  \href{http://dx.doi.org/10.1016/j.physletb.2021.136111}{\emph{Phys. Lett. B}
  {\bf 814} (2021) 136111}, [\href{https://arxiv.org/abs/2007.09420}{{\tt
  2007.09420}}].

\bibitem{Buchel:2020jfs}
A.~Buchel, \emph{{Fate of the conformal order}},
  \href{http://dx.doi.org/10.1103/PhysRevD.103.026008}{\emph{Phys. Rev. D} {\bf
  103} (2021) 026008}, [\href{https://arxiv.org/abs/2011.11509}{{\tt
  2011.11509}}].

\bibitem{Buchel:2022zxl}
A.~Buchel, \emph{{The quest for a conifold conformal order}},
  \href{http://dx.doi.org/10.1007/JHEP08(2022)080}{\emph{JHEP} {\bf 08} (2022)
  080}, [\href{https://arxiv.org/abs/2205.00612}{{\tt 2205.00612}}].

\bibitem{Buchel:2008ae}
A.~Buchel, R.~C. Myers, M.~F. Paulos and A.~Sinha, \emph{{Universal holographic
  hydrodynamics at finite coupling}},
  \href{http://dx.doi.org/10.1016/j.physletb.2008.10.003}{\emph{Phys. Lett. B}
  {\bf 669} (2008) 364--370}, [\href{https://arxiv.org/abs/0808.1837}{{\tt
  0808.1837}}].

\bibitem{Baier:2007ix}
R.~Baier, P.~Romatschke, D.~T. Son, A.~O. Starinets and M.~A. Stephanov,
  \emph{{Relativistic viscous hydrodynamics, conformal invariance, and
  holography}},
  \href{http://dx.doi.org/10.1088/1126-6708/2008/04/100}{\emph{JHEP} {\bf 04}
  (2008) 100}, [\href{https://arxiv.org/abs/0712.2451}{{\tt 0712.2451}}].

\bibitem{Bhattacharyya:2007vjd}
S.~Bhattacharyya, V.~E. Hubeny, S.~Minwalla and M.~Rangamani, \emph{{Nonlinear
  Fluid Dynamics from Gravity}},
  \href{http://dx.doi.org/10.1088/1126-6708/2008/02/045}{\emph{JHEP} {\bf 02}
  (2008) 045}, [\href{https://arxiv.org/abs/0712.2456}{{\tt 0712.2456}}].

\bibitem{Buchel:2009hv}
A.~Buchel, \emph{{Relaxation time of non-conformal plasma}},
  \href{http://dx.doi.org/10.1016/j.physletb.2009.10.007}{\emph{Phys. Lett. B}
  {\bf 681} (2009) 200--203}, [\href{https://arxiv.org/abs/0908.0108}{{\tt
  0908.0108}}].

\bibitem{Buchel:2008bz}
A.~Buchel and M.~Paulos, \emph{{Relaxation time of a CFT plasma at finite
  coupling}},
  \href{http://dx.doi.org/10.1016/j.nuclphysb.2008.07.002}{\emph{Nucl. Phys. B}
  {\bf 805} (2008) 59--71}, [\href{https://arxiv.org/abs/0806.0788}{{\tt
  0806.0788}}].

\bibitem{Buchel:2008kd}
A.~Buchel and M.~Paulos, \emph{{Second order hydrodynamics of a CFT plasma from
  boost invariant expansion}},
  \href{http://dx.doi.org/10.1016/j.nuclphysb.2008.10.012}{\emph{Nucl. Phys. B}
  {\bf 810} (2009) 40--65}, [\href{https://arxiv.org/abs/0808.1601}{{\tt
  0808.1601}}].

\bibitem{Saremi:2011nh}
O.~Saremi and K.~A. Sohrabi, \emph{{Causal three-point functions and nonlinear
  second-order hydrodynamic coefficients in AdS/CFT}},
  \href{http://dx.doi.org/10.1007/JHEP11(2011)147}{\emph{JHEP} {\bf 11} (2011)
  147}, [\href{https://arxiv.org/abs/1105.4870}{{\tt 1105.4870}}].

\bibitem{Grozdanov:2016fkt}
S.~Grozdanov and A.~O. Starinets, \emph{{Second-order transport, quasinormal
  modes and zero-viscosity limit in the Gauss-Bonnet holographic fluid}},
  \href{http://dx.doi.org/10.1007/JHEP03(2017)166}{\emph{JHEP} {\bf 03} (2017)
  166}, [\href{https://arxiv.org/abs/1611.07053}{{\tt 1611.07053}}].

\bibitem{Witten:1998zw}
E.~Witten, \emph{{Anti-de Sitter space, thermal phase transition, and
  confinement in gauge theories}},
  \href{http://dx.doi.org/10.4310/ATMP.1998.v2.n3.a3}{\emph{Adv. Theor. Math.
  Phys.} {\bf 2} (1998) 505--532},
  [\href{https://arxiv.org/abs/hep-th/9803131}{{\tt hep-th/9803131}}].

\bibitem{Wald:1993nt}
R.~M. Wald, \emph{{Black hole entropy is the Noether charge}},
  \href{http://dx.doi.org/10.1103/PhysRevD.48.R3427}{\emph{Phys. Rev. D} {\bf
  48} (1993) R3427--R3431}, [\href{https://arxiv.org/abs/gr-qc/9307038}{{\tt
  gr-qc/9307038}}].

\bibitem{Maldacena:1997re}
J.~M. Maldacena, \emph{{The Large N limit of superconformal field theories and
  supergravity}}, \href{http://dx.doi.org/10.1023/A:1026654312961,
  10.4310/ATMP.1998.v2.n2.a1}{\emph{Int. J. Theor. Phys.} {\bf 38} (1999)
  1113--1133}, [\href{https://arxiv.org/abs/hep-th/9711200}{{\tt
  hep-th/9711200}}].

\bibitem{Aharony:1999ti}
O.~Aharony, S.~S. Gubser, J.~M. Maldacena, H.~Ooguri and Y.~Oz, \emph{{Large N
  field theories, string theory and gravity}},
  \href{http://dx.doi.org/10.1016/S0370-1573(99)00083-6}{\emph{Phys. Rept.}
  {\bf 323} (2000) 183--386}, [\href{https://arxiv.org/abs/hep-th/9905111}{{\tt
  hep-th/9905111}}].

\bibitem{Skenderis:2002wp}
K.~Skenderis, \emph{{Lecture notes on holographic renormalization}},
  \href{http://dx.doi.org/10.1088/0264-9381/19/22/306}{\emph{Class. Quant.
  Grav.} {\bf 19} (2002) 5849--5876},
  [\href{https://arxiv.org/abs/hep-th/0209067}{{\tt hep-th/0209067}}].

\bibitem{Buchel:2004hw}
A.~Buchel, \emph{{N=2* hydrodynamics}},
  \href{http://dx.doi.org/10.1016/j.nuclphysb.2004.11.039}{\emph{Nucl. Phys.}
  {\bf B708} (2005) 451--466},
  [\href{https://arxiv.org/abs/hep-th/0406200}{{\tt hep-th/0406200}}].

\bibitem{Policastro:2002se}
G.~Policastro, D.~T. Son and A.~O. Starinets, \emph{{From AdS / CFT
  correspondence to hydrodynamics}},
  \href{http://dx.doi.org/10.1088/1126-6708/2002/09/043}{\emph{JHEP} {\bf 09}
  (2002) 043}, [\href{https://arxiv.org/abs/hep-th/0205052}{{\tt
  hep-th/0205052}}].

\bibitem{Son:2002sd}
D.~T. Son and A.~O. Starinets, \emph{{Minkowski space correlators in AdS / CFT
  correspondence: Recipe and applications}},
  \href{http://dx.doi.org/10.1088/1126-6708/2002/09/042}{\emph{JHEP} {\bf 09}
  (2002) 042}, [\href{https://arxiv.org/abs/hep-th/0205051}{{\tt
  hep-th/0205051}}].

\bibitem{Cremonini:2009ih}
S.~Cremonini, J.~T. Liu and P.~Szepietowski, \emph{{Higher Derivative
  Corrections to R-charged Black Holes: Boundary Counterterms and the
  Mass-Charge Relation}},
  \href{http://dx.doi.org/10.1007/JHEP03(2010)042}{\emph{JHEP} {\bf 03} (2010)
  042}, [\href{https://arxiv.org/abs/0910.5159}{{\tt 0910.5159}}].

\bibitem{Buchel:2004qq}
A.~Buchel, \emph{{On universality of stress-energy tensor correlation functions
  in supergravity}},
  \href{http://dx.doi.org/10.1016/j.physletb.2005.01.052}{\emph{Phys. Lett. B}
  {\bf 609} (2005) 392--401}, [\href{https://arxiv.org/abs/hep-th/0408095}{{\tt
  hep-th/0408095}}].

\end{thebibliography}\endgroup

\end{document}